\documentclass[oneside,12pt]{book}
\usepackage{epsf,epsfig,graphics}
\begin{document}
\frontmatter
\pagestyle{plain}

\thispagestyle{empty}

\begin{center}
{\Large\bf
\centerline{Nonperturbative Lattice Simulation of}
\centerline{High Multiplicity Cross Section Bound in $\phi^4_3$ on} 
\centerline{Beowulf Supercomputer}
}
\vskip 1cm
Yeo-Yie Charng \\
\vskip 1cm
{\it Department of Physics \\
University of Pittsburgh\\
Pittsburgh,PA 15260}
\end{center}
\vskip 1cm
\centerline{\bf Abstract}

  In this thesis, we have investigated the possibility of large cross sections 
at large 
multiplicity in weakly coupled three dimensional $\phi^4$ theory using 
Monte Carlo Simulation methods. We have built a Beowulf Supercomputer
for this purpose. We use spectral function sum rules to derive a bound on the
total cross section where the quantity determining the bound can be
measured by Monte Carlo simulation in Euclidean space. We determine the 
critical threshold energy for large high multiplicity cross section 
according to the analysis of  
M.B. Volosion and E.N. Argyres, R.M.P. Kleiss, and C.G. Papadopoulos.
We compare the simulation results with the perturbation results
and see no evidence for large cross section in the range where tree 
diagram estimates suggest they should exist.

\tableofcontents

\listoffigures

\listoftables

\pagestyle{myheadings}
\markright{}

\mainmatter

\chapter*{Overview}

  Chapter 1 of this thesis is the introduction to the high multiplicity
cross section problem and the idea to solve the problem using Monte
Carlo simulation. Chapter 2 describes the detail of design and 
construction of the Beowulf parallel supercomputer. In chapter 3, 
the sum rules and the bound for the cross section are derived.

  In chapter 4, a perturbative  estimate of the critical energy is given. 
And the relationship between the critical
energy and the Euclidean momentum on the lattice is determined. 
In chapter 5, the Monte Carlo simulation is described. 
The scaling behavior and continuum limit are also discussed. 
In chapter 6, I will use the perturbation method to calculate the cross
sections.  
Chapter 7 presents the results from the simulation. 
Finally, I will compare the 
results with the perturbative calculations and discuss conclusions. 
\chapter{Introduction}

   In 1990, Ringwald \cite{Ri:nphb350} and Espinosa \cite{Es:nphb343} presented 
a calculation of possible 
instanton induced baryon number violation accompanied by enhanced production 
of large number of Higgs and W,Z bosons at high energy($E \sim n/g$,where $n$
 is the number of out going particles and $g$ is a generic coupling constant). 
This  has called attention to a more general and profound problem of
quantum field theory. The problem is: for a given renormalized quantum field 
theory with weak renormalized coupling constant, what is the nature of the
multi-particle production in that theory. One can think two extreme cases. 
In the first case, 
the cross section is perturbatively small[$\sim g^n \times O(1)$]. 
In the second case, when the number of out going particles $n \sim 1/g$, the
multiparticle production cross section becomes unsuppressed and the inclusive
cross sections saturate unitarity bounds.

  It has long been known that the perturbation-theory expansion in a theory 
with weak coupling fails in high orders because of the factorial growth of
the coefficients in the series \cite{Zj:phyrp70}. This problem can 
not be solved 
perturbatively because the perturbation series  diverges
rapidly at high order. 
In fact, independent of the instanton calculations,
Goldberg \cite{Go:phyletb246}, Cornwall\cite{Co:phyletb243} and, later,  
Voloshin\cite{Vo:phyletb293}, Argyres, Kleiss, and Papadopoulos
\cite{AKG:nphb341}
point out that in $\phi^4$ scalar field theory the contributions of just the 
tree graphs to the multiparticle production amplitudes gives the $n!g^n$ 
behavior. This suggests the possibility of the second case.

  Here we have a different approach to solve this nonperturbative problem.  
We propose to use Monte Carlo simulation on a lattice to 
determine an upper bound on the cross section of the multiparticle 
production in $\phi^4_3$ theory. The original application of this approach was 
to two dimensional $\phi^4_2$ theory\cite{Wm:prl}. However, the case of $1+1$ 
dimensions is a very special case in field theory.(no angular momentum...etc)
Although the invesgation in $1+1$ dimension showed no signal of large 
multiparticle production, the result
might be special for the $1+1$ dimension case. In this research,
we will use Monte Carlo simulation to establish a bound on the multiparticle
cross section in three dimensions($1+2$ dimensions in Minkowski).   
The tree diagram considerations which suggest large $1\rightarrow n$ 
amplitudes in 
$\phi^4$ are mainly combinatoric, and hold as well in three dimensions
as in two or four. Working in three dimensions has several advantages.
First, we can work on large $N^3$ lattices as long as our computers can 
provide the memory and give us reasonable performance. 
Second, three dimensional 
$\phi^4$ theory is a superrenormalizable theory. 
The wave function renormalization constant $Z$ is finite in the $\phi^4_3$
theory. The other important issue in our simulation is that we can do
the simulation in three dimensional Euclidean space. The constructive quantum 
field theorists have proved that the $\phi^4_3$ field theory exists as the 
analytic continuation of the continuum limit of the Euclidean lattice theory
\cite{GJ:qphy,FF:rwqft}.  In four dimensions,
nonperturbatively, we would have to deal with the presumed triviality of the
theory.

  Even though we worked on three dimensional Euclidean lattices rather than 
four, the simulation itself is still a huge computing task. 
This is because the many degree of freedoms required for sensitivity to 
multiparticle production requires large lattice. For this reason,
we will still need a supercomputer to do the job. To solve this problem,
we built a Beowulf parallel supercomputer with sixteen CPUs. We also 
use the Message Passing Interface(MPI) function call to parallelize our 
simulation
program. We are able to achieve overall 7.4 Gflops at High Performance 
Computing Linpack Benchmark(HPL). The theoretical peak performance for sixteen
CPUs is 12 Gflops. Compare with 16 Gflops CRAY C90 at 16 CPUs, our parallel
cluster has a much better price/performance ratio.

\chapter{Beowulf Parallel Supercomputer}

\section{The Beowulf Project History}

The Beowulf Project was started by Donald Becker when he moved to CESDIS in 
early 1994. CESDIS was located at NASA's Goddard Space Flight Center, and 
was operated for NASA by USRA\cite{BW:www}. 

In the summer of 1994 the first Beowulf 16 node demonstration cluster was 
constructed for the Earth and Space Sciences project, (ESS). The project 
quickly spread to other NASA sites, other R\&D labs and to universities around 
the world. The project's scope and the number of Beowulf installations have 
grown over the years. Today, the fastest Beowulf cluster can reaches 143 Gflops
(HPL benchmark) with 528 Pentium III 800 CPUs which ranks at 126 in the top500 
supercomputer list in Nov, 2000\cite{Tp:500}. With more and faster CPUs and 
even faster 
network links, Beowulf supercomputer can compete with any commercial high 
performance supercomputer such as SGI Origin 2000 or Cray T3E and 
costs much less.

\section{Definition of the Beowulf Supercomputer}

Beowulf is a multi computer architecture which can be used for parallel 
computations.  It is a system which usually consists of one server node, and 
one or more client nodes connected together via Ethernet or some other network.
It is a system built using commodity hardware components, like any PC capable 
of running Linux, standard Ethernet adapters, and switches. It does not 
contain any custom hardware components and is trivially reproducible. Beowulf 
also uses commodity software like the Linux or FreeBSD operating system, 
Parallel Virtual Machine (PVM) and Message Passing Interface (MPI). The server 
node controls the whole cluster and serves files to the client nodes.  It 
might also be the cluster's console and gateway to the outside world if there 
is no particular annex machine. Large Beowulf machines 
might have more than one server node, and possibly other nodes
dedicated to particular tasks, for example consoles or monitoring stations. 
The client nodes in the cluster don't sit on people's desks; they are 
dedicated to running cluster jobs only.
In most cases client nodes in a Beowulf system are dumb and
are configured and controlled by the server node, and do only what they 
are told to do. In a disk-less client configuration, client nodes don't even 
know their IP address or name until the server tells them what it is. One of 
the main differences between Beowulf and a Cluster of Workstations (COW) is 
the fact that Beowulf behaves more like a single machine rather than many 
workstations. In most cases client nodes do not have keyboards or monitors, 
and are accessed only via remote login or possibly serial terminal. Beowulf 
client nodes can be thought of as a CPU + memory package which can be plugged 
in to the cluster, just like a CPU or memory module can be plugged into a 
motherboard. In general, the user can only login to the front end and submit 
their
jobs to the cluster. The queueing system will receive the jobs and run them 
in different nodes\cite{BW:ht}.

   To get better performance,  one can also use some dedicated parts which can 
not be purchased on the commodity market. This is true especially when 
people need a very fast link between the nodes. One can use Giga bits 
Ethernet or Myrinet(240MB/s at 64-bit, 33MHz PCI) or even 
Quadric(340MB/s) system for the fast 
link. The other way to improve the floating point performance is to use DSP 
chips such as Texas Instrument TMS320C6xxx. These chips can boost the floating
point performance up to 3 Gflops. However, the drawback is that they are 
expensive and some of them lack driver support.

\section{Architecture and System Design}

\begin{figure}
\begin{center}

\vskip 1cm
\leavevmode
\epsfxsize 6in
\epsfbox{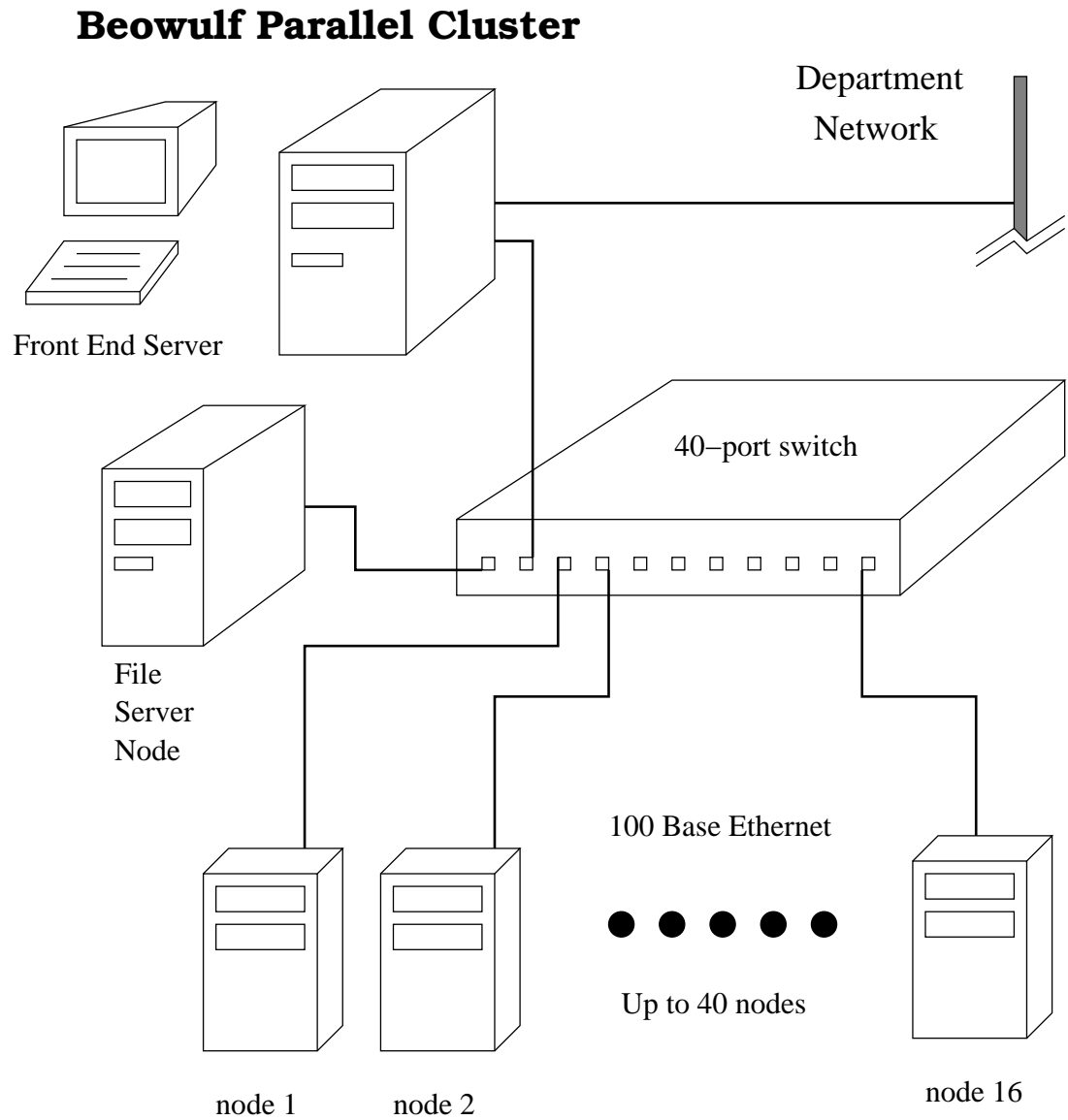}
\vskip 1cm

\caption{The architecture of the Beowulf Supercomputer}
\label{fig:cluster}
\end{center}
\end{figure}

  Our Beowulf Supercomputer is designed to solve large-scale scientific 
problems.
Because we need to run our simulation on a large lattices, we also need huge 
amount of memory. Each node within the cluster is running the freely 
available Linux operating system. Parallel capabilities are achieved through 
the use of the MPI and PVM communication protocols. Job scheduling and 
execution for all users is handled by the Distributed Queueing System(DQS).
The system contain a server node, a annex node, and 16 client nodes. All of 
these machines are networked by 100 BaseT Ethernet with low latency switch.
(see Figure \ref{fig:cluster}) This architecture is so call distributed 
memory(or local memory) 
architecture which is good to run both Single Program, Multiple Data(SPMD) 
and Multiple Instruction, Multiple Data(MIMD) program.
The machine was assembled entirely from individually chosen 
commodity hardware components. With the exception of the Ethernet switch, 
all of the computer hardware within the cluster was built from the component 
level. This low-level approach allowed the developers to optimize the overall
performance by providing us the freedom to choose the fastest and most reliable 
components to fit our budget.  Likewise, all matters of OS and software package 
configuration were carried out in-house in an effort to achieve optimal 
settings for speed and stability\cite{Tp:www}.

\subsection{Hardware Configuration}

The server node contains:
\begin{itemize}
\item Pentium III 500 MHz CPU 
\item ASUS P2B-S Motherboard
\item 384 MB Samsung SDRAM 
\item 27 GB on two Ultra2 Wide LVD Quantum Atlas III SCSI HDs 
\item Seagate 12/24 GB DAT Tape Backup System 
\item Kingston KNE-100TX Tulip-based Ethernet card 
\item PCI video card 
\item Linksys I/O console selector switch 
\end{itemize}

Each client nodes has:
\begin{itemize}
\item Pentium III 744 MHz Coppermine CPU 
\item ASUS P2B-F Motherboard 
\item 768 MB Micron PC133 SDRAM 
\item 4.3 GB UDMA Quantum Fireball HD 
\item Kingston KNE-100TX Tulip-based Ethernet card 
\item PCI video card (for boot-up only)
\end{itemize}

The annex machine is the gateway of the cluster. It has:
\begin{itemize}
\item ABIT BP6 Dual Celeron Motherboard with Two 450MHz CPUs 
\item 256 MB Samsung SDRAM 
\item 6 GB UDMA33 IDE HD 
\item Kingston KNE-100TX Tulip-based Ethernet card 
\item 3Com 10base-T Ethernet card (for outside LAN)
\item PCI video card 
\end{itemize}

\subsection{Software Configuration}

  All of the nodes are running Linux Operating System. The Clients nodes share
the user and utility files with the other via NFS(network file system). 
For the message passing protocol, we use message passing Interface(MPI), 
Parallel Virtual Machine(PVM), High Performance Fortran(HPF), and newly
developed Cactus Code. The DQS queueing system can handle both serial and 
parallel jobs.

\section{Performance and Benchmark}

  The performance of our Beowulf Cluster is quite remarkable. We tested it with
two different parallel benchmark program which are recommended by the top500 
cluster web site, NAS and HPL\cite{Tp:500bm}.  
The NAS Parallel Benchmarks (NPB) are a set of 8 programs designed to help 
evaluate the performance of parallel supercomputers. The benchmarks, which are 
derived from computational fluid dynamics (CFD)
applications, consist of five kernels and three pseudo-applications. 

The detail description for each benchmark routine is following:

\begin{itemize}
\item    CG - solution of a structured sparse linear system by the conjugate gradient method 
\item    FT - this benchmark contains the computational kernel of a three dimensional FFT-based spectral
    method 
\item    MG - Multigrid Benchmark uses a multigrid method to compute the solution of the three-dimensional
    scalar Poisson equation 
\item    LU - a simulated CFD application which uses symmetric successive over-relaxation (SSOR) to solve a
    block lower triangular-block upper triangular system of equations resulting from an unfactored implicit
    finite-difference discretization of the Navier-Stokes equations in three dimensions 
\item    IS - parallel sort over small integers 
\item    EP - embarrassingly parallel benchmark 
\item    SP and BT - SP and BT are simulated CFD applications that solve 
systems of equations resulting from an approximately factored implicit 
finite-difference discretization of the Navier-Stokes equations. 
\item    BT code solves block-tridiagonal systems of 5x5 blocks; the SP code solves scalar pentadiagonal systems
    resulting from full diagonalization of the approximately factored scheme. 
\end{itemize}

   Here we show the  class C(large size) results in Table 
\ref{table:NAS}. Compare with CRAY T3E 1200, 
one can see that the Beowulf Supercomputer is even faster then T3E 1200 at 
same number of CPUs. However, for a heavy communication job, the T3E will
be faster because of the faster interlink between the CPUs.

\begin{table}[t]

\caption{NAS Benchmark} \vskip 24pt
\label{table:NAS}
\begin{center}
\begin{tabular}{|c|c|c|c|}
\hline
Benchmark & Number of & Beowulf & CRAY T3E1200 \\
Name & Nodes &  & (from NAS,NASA) \\
\hline
BT class=C & 16 & 1250.88 Mflops & 1088.3 Mflops\\
\hline
IS class=C & 16 & 28.83 Mflops & 26.9 Mflops\\
\hline
CG class=C & 16 & 211.63 Mflops & 148.7 Mflops \\
\hline
EP class=C & 16 & 52.56 Mflops & 54.8 Mflops \\
\hline
LU class=C & 16 & 1380.13 Mflops & 1169.5 Mflops\\
\hline
MG class=C & 16 & 1071.59 Mflops & 1616.8 Mflops \\
\hline
SP class=C & 16 & 871.19 Mflops & 800.4 Mflops \\
\hline
FT class=C & 16 & 392.64 Mflops & NA \\
\hline
\end{tabular}
\end{center}
\end{table}

\begin{figure}[b]
\begin{center}
\leavevmode
\epsfxsize 5in
\epsfbox{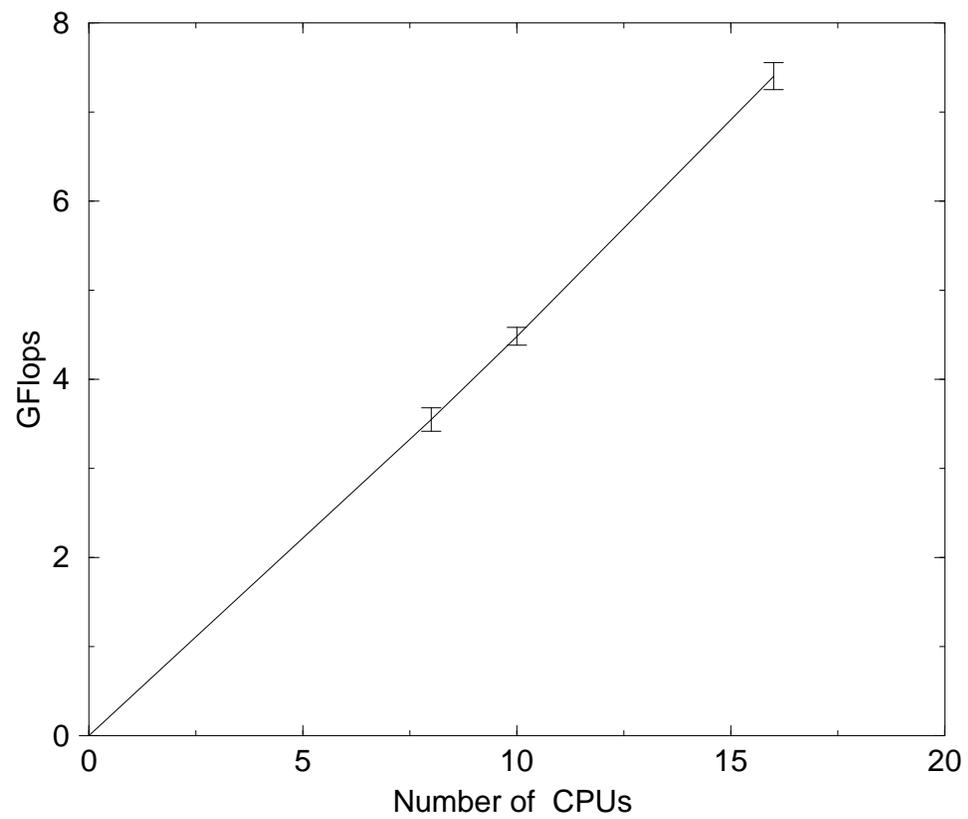}
\caption{The HPL Benchmark shows the speed scales up linearly.}
\label{figure:ben}
\end{center}
\end{figure}

   The second benchmark is High Performance Linpack. This benchmark is the one
used in the Top500 supercomputer list\cite{Tp:500}. HPL is a software package 
that 
solves a (random) dense linear system  in   double  precision  (64   bits)   
arithmetic   on distributed-memory  computers.   It can thus be regarded as a
portable as well as  freely  available implementation  of the High Performance 
Computing Linpack Benchmark.  Here we run three different sizes of the problem
with different number of CPUs. The table \ref{table:ben} shows the results. 
One can see
the fastest result is 7.4 Gflops at 16 CPUs($4\times 4$).  
The Figure \ref{figure:ben} shows that up to 16 
CPUs, the speed scales up linearly. At the scale of 32 nodes or 
above\cite{BW:www}, the
performance saturates the bandwidth of the 100Mbits network. The Giga bits
network can boost the performance  more. However, the Giga bits 
network suffers from the TCP/IP protocol which was designed for 10Mbits network
a longtime ago. For the Giga bits Ethernet link, one can only optimize either 
for the bandwidth or for the latency.

\begin{table}[h]
\caption{HPL Benchmark} \vskip 24pt
\label{table:ben}
\begin{center}
\begin{tabular}{|c|c|c|c|c|c|}
\hline
Matrix Size & Block size & P & Q & Time & Results(Gflops) \\
\hline
20000 & 200 & 2 & 4 & 1501 sec & 3.55 \\
\hline
25000 & 200 & 2 & 5 & 2324 sec & 4.48 \\
\hline
34000 & 200 & 4 & 4 & 3545 sec & 7.4 \\
\hline
\end{tabular}
\end{center}
\end{table}

\chapter{The Cross Section and Bound}

   In this chapter, we want to determine the upper bound for the inclusive
cross section of the multiparticle production. We will derive the bound
in four steps in following four sections. First of all, we will specify the
scattering process and define the inclusive cross section. Secondly, use the
spectral representation of the two-point function and find the relation 
between the inclusive cross
section and the spectral function.  Thirdly, define $G^{-1}$, $Z'$, and $m'$ 
and give the spectral representation of the $G^{-1}$ function. In the last
section, we will find the upper bound of the inclusive cross 
section.

\section{Definition of  the inclusive cross section}

\begin{figure}[h]
\begin{center}
\leavevmode
\epsfxsize 4in
\epsfbox{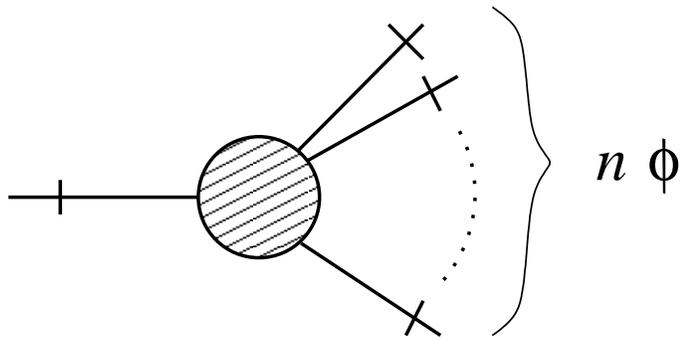}
\caption{The amplitude $A(n;E)$ for $\phi \rightarrow n\phi$.}
\label{fig:feyn1ton}
\end{center}
\end{figure}

\begin{figure}[h]
\begin{center}
\leavevmode
\epsfxsize 5in
\epsfbox{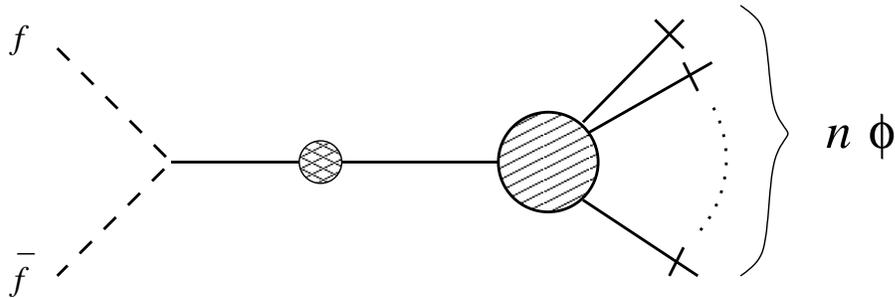}
\caption{The amplitude for $f\bar{f} \rightarrow n\phi$, to lowest order in $g_y$}
\label{fig:fftoblob}
\end{center}
\end{figure}

  As we mentioned in chapter one, we want to study  the amplitudes for one 
off-shell(time-like) $\phi$  particle to go to $n$ on-shell $\phi$ particles
(see Figure \ref{fig:feyn1ton}).
There are the amplitudes for which extensive tree diagram analysis have been 
done\cite{Go:phyletb246,Co:phyletb243,Vo:phyletb293,AKG:nphb341,Br:prd46}. 
These amplitudes may be considered as ``physical" if we introduce a
weak Yukawa coupling of the $\phi$ field to a fermion
\begin{equation}
{\cal L}_y = g_y\bar{\psi}\psi\phi,
\end{equation}
and compute the cross section for $f\bar{f}$ to annihilate into $n\phi$ 
particles.(See Figure \ref{fig:fftoblob}) 

This is similar to the $e^+e^-$ annihilation into hardrons in QCD.
The transition probability for Figure \ref{fig:feyn1ton} is
\begin{equation} 
m^2\Gamma(n,E) = \frac{1}{n!}\int d\Phi_n(E)|A(n;E)|^2  \label{eq:transition}
\end{equation}
where $E=\sqrt{s}$. The amplitude $A(n;E)$ is
\begin{equation} 
A(n;E) = \underline{G}_{(n+1)}(p_1,p_2,\cdots,p_n)\hspace{-3.7cm}/ \hskip 2cm   
\end{equation}
which is the connected $(n+1)$ point Green function.
The underscore means amputated Green function (no external line factors) and 
the slash means renormalized Green function (no $Z$ factions in the 
Lehmann-Symanzik-Zimmermann reduction formulas). It is convenient define a 
partial ``cross section" of Figure \ref{fig:feyn1ton} as a kinematic factor 
times equation \ref{eq:transition}
\begin{equation}
\sigma_n(s) = \frac{m^2}{2\pi s^2}\Gamma(n;E)
\end{equation}
for $n$ out going particles. The the cross section of $f\bar{f} 
\rightarrow n\phi$ in three dimensions then can be written as(to lowest 
order in Yukawa coupling 
but to all orders in the $\phi$ self-coupling\cite{AKG:nphb341})

\begin{equation}
\sigma_{f\bar{f} \rightarrow n\phi}(s) = \frac{3\pi g_y^2}{8}|sG\hspace{-0.25cm}/_M|^2 \times
\sigma_n(s)
\end{equation}
Finally, we include all possible number of out going particles and define the
{\em inclusive cross section} as
\begin{eqnarray}
\sigma(s) & = & \sum^{\infty}_{n} \sigma_n(s) \\
& = & \frac{1}{2\pi s^2}\sum_n \frac{1}{n!}
\int d\Phi_n |\underline{G}\hspace{-3mm}/~_{(n+1)}(p_1,\cdots,p_n)|^2 \label{in:cro}
\end{eqnarray}

\section{Spectral Representation and Sum Rules}

  In the last section, we defined the inclusive cross section for the 
multiparticle
final state. One can relate the inclusive cross section to the spectral 
function of the two point function. Let $\phi_0(x)$ be the canonical, 
unrenormalized scalar field in Minkowski space. We have the Wightman two point
function
\begin{equation}
W(x,y) = \langle \Omega|\phi(x)\phi(y)|\Omega \rangle
\end{equation}
The two point Green function(or $\tau$ function) is the time ordered 
Wightman function
\begin{equation}
G(x,y) = \langle \Omega|T(\phi(x)\phi(y)|\Omega \rangle
\end{equation}
Define the spectrum of physical state in three dimensions as
\begin{eqnarray}
{\bf 1} & = & |\Omega\rangle\langle\Omega| + 
\int\frac{dp^2}{(2\pi)^22E_p} |p\rangle\langle p| \nonumber\\
&&+ \sum^{\infty}_{n=2}\frac1{n!}\int\prod\frac{d^2p}{(2\pi)^22E_n} 
|p_1,\cdots,p_n\rangle\langle p_1,\cdots,p_n|\cdots \nonumber \\
& = & \sum^{\infty}_{n} |n\rangle\langle n| \label{in:phs}
\end{eqnarray}
and the phase space integral as
\begin{equation}
\int d\Phi_n = (2\pi)^3\delta^3(p_1+\cdots+p_n)\frac1{n!}\prod^n_{a=1}
\int\frac{d^2p_a}{(2\pi)^22E_a}
\end{equation}
The Wightman two point function has the spectral representation
\begin{eqnarray}
\langle\Omega|\phi(x)\phi(y)|\Omega\rangle & = & \int \frac{d^3q}{(2\pi)^3} \theta(q_0)
e^{-iq(x-y)}\rho(q^2) \label{tp:rho}\\
& = &  \int^{\infty}_0 d\kappa^2 \rho(\kappa^2)\Delta^{(+)}(x-y,\kappa^2)
\end{eqnarray}
and the Green function is
\begin{eqnarray}
G(x-y) & = & \int^{\infty}_{0}d\mu^2 \rho(\mu^2)\Delta_F(x-y;\mu^2) 
\label{G:eucd}\nonumber\\
& = & Z\Delta_F(x-y;m^2) + \int^{\infty}_{4m^2} d\mu^2 
\rho(\mu^2)\Delta_F(x-y;\mu^2)
\end{eqnarray}
where the spectral function $\rho(\mu^2)$ in three dimensions is
\begin{equation}
\theta(p^0)\rho(p^2) = (2\pi)^2\sum^{\infty}_{n} 
\delta^3(p-p_1-p_2-\cdots-p_n)|\langle\Omega|\phi(0)|n\rangle |^2
\end{equation}
Fourier Transform equation \ref{G:eucd} and make the analytic continuation to 
Euclidean space.
\begin{eqnarray}
G_M(p^2_M) \rightarrow
G(p^2) &=& \int d\kappa^2 \frac{\rho(\kappa^2)}{p^2+\kappa^2}\nonumber\\
&=& \frac{Z}{p^2+m^2}+\int d\kappa^2 \frac{\hat{\rho}(\kappa^2)}{p^2+\kappa^2}
\label{tp:fn}\\
&=& ZG(p^2)\hspace{-.95cm}/
\end{eqnarray}
The $\hat{\rho}$ is the multiparticle(n $\geq$ 2) contribution to the spectral 
function.
In three dimensions, the field strength renormalization is finite and the 
canonical sum rule holds:
\begin{equation}
1 = \int^{\infty}_{0} d\kappa^2\rho(\kappa^2) = Z + 
\int d\kappa^2 \hat{\rho}(\kappa^2). \label{eq:spctrl}
\end{equation}
Positivity of the spectral function implies
\begin{equation}
0 \leq Z \leq 1
\end{equation}
Note that the equation \ref{tp:rho} is just the Fourier transform of 
$\rho(q^2)$, so the
inverse Fourier transform gives(choose $y=0$)
\begin{equation}
2\pi\rho(q^2) = \int d^3x e^{iqx}\langle\Omega|\phi(x)\phi(0)|\Omega\rangle
\end{equation}
Introduce a complete set of intermediate states \ref{in:phs} and apply the 
LSZ reduction formula to find
\begin{equation}
2\pi\rho(q^2) = G_M(q)F(q)G_M(q) \label{eq:rho}
\end{equation}
where
\begin{equation}
F(q) = \sum^{\infty}_n \frac{Z^n}{n!}\int d\Phi_n |\underline{G}\hspace{-3mm}/~^c_{(n+1)}(p_1,\cdots,p_n,q)|^2 \label{eq:Fq}
\end{equation}
Here we only 
consider the connected part of the Green function. We can use equation \ref{eq:rho}, \ref{eq:Fq}, and \ref{in:cro} to get the relation between $\sigma$ and 
$\rho$.  Define the renormalized quantities as
\begin{equation}
G_0 = ZG\hspace{-3mm}/ ~,~~~~~~ \rho = Z\rho\hspace{-2mm}/, ~~~~~~
G_{0_{(n+1)}} = Z^{\frac{n+1}{2}}G\hspace{-3mm}/~_{(n+1)} \label{rn:cnd}
\end{equation}
and using these equations, we find
\begin{equation}
\sigma(s) = \frac{\rho\hspace{-2mm}/~(s)}{|sG\hspace{-3mm}/~_M|^2} 
\label{cr:den}
\end{equation}

\section{Spectral Representation of $G^{-1}$}

   From equation \ref{cr:den}, one can see that the inclusive cross section is 
closely related to the spectral function and the inverse of the 
Green function.  The spectral representation of $G_M(P)$ in Minkowski space is
\begin{equation}
G_M(P) = \frac{Z}{P^2-m^2} + \int^{\infty}_{4m^2} d\mu^2 
\frac{\hat{\rho}(\mu^2)}{p^2-\mu^2+i\epsilon} 
\label{eq:sptrl}
\end{equation}

\begin{figure}[tbp]
\begin{center}
\epsfxsize= 5.0 in
\epsfbox{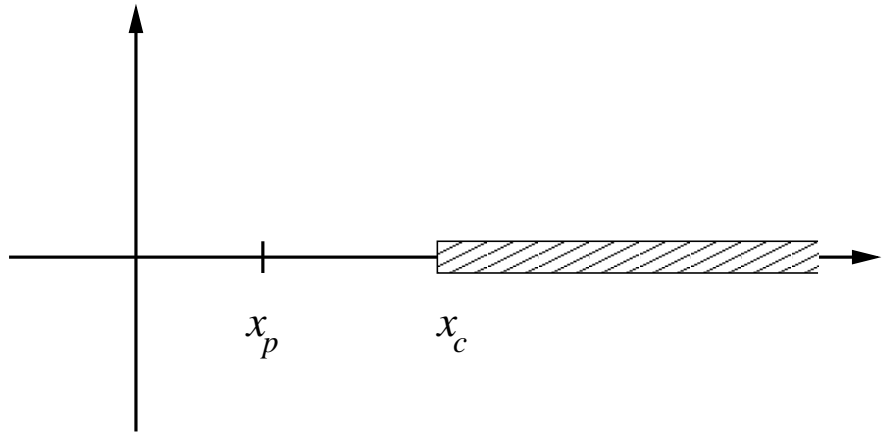}
\caption{The pole and structure for $g(x)$}
\label{figure:polecut}
\end{center}
\end{figure}

Consider a simple complex function with the same pole and cut 
structure as
\begin{equation}
g(z)=\frac{Z}{z-x_p} + \int^{\infty}_{x_c}dx'\frac{\hat{\rho}(x')}{z-x'}
\label{eq:gz}
\end{equation}
The function has a pole at $x_p$ and cut at $x>x_c$. The pole and cut 
structure are shown in Figure \ref{figure:polecut}.
To analyze the pole
structure of $g^{-1}(z)$, we want to know the zero structure of $g(z)$. 
The positivity of $\rho$ implies $g(z)$ has no complex zero and no zero 
when $x < x_p$ or $x > x_c$. The only possible zero for $g(\xi)$ is 
when $x_p < \xi < x_c$.

Because $g(z)g^{-1}(z)=1$, $g^{-1}(z)$ should have the same cut structure
as $g(z)$. We can write
\begin{equation}
g^{-1}(z)= f(z)+\sum_i\frac{{\cal C}_i}{z-\xi_i} + \int^{\infty}_{x_c} dx'
\frac{\gamma(x')}{x'-z} \label{eq:gz-1}
\end{equation}
where ${\cal C}_i$ is a positive constant and $\gamma(x)$ is the spectral
function for $g^{-1}(x)$
and $f(z)$ is just a polynomial function. We can simplify the calculation
by considering the case $g(z)$ has no zero(${\cal C}_i=0$). The multiple 
zero case will be discussed in the Appendix. So the equation \ref{eq:gz-1}
becomes
\begin{equation}
g^{-1}(z)= f(z)+ \int^{\infty}_{x_c} dx'
\frac{\gamma(x')}{x'-z} \label{eq:gz-1nozero}
\end{equation}

Because the pole of $g(z)$ is the zero of the $g^{-1}(z)$, one can
impose the condition $g^{-1}(x_p)=0$ and $g(z)g^{-1}(z)=1$ into equation
\ref{eq:gz},\ref{eq:gz-1nozero} and get
\begin{equation}
g^{-1}(z)=(z-x_p)\left( 1+ \int^{\infty}_{x_c}
dx' \frac{\gamma(x')}{(x'-z)(x'-x_p)} \right) \label{gz:inv}
\end{equation}

With the $i\epsilon$ prescription of equation \ref{eq:sptrl}, we can calculate
the imaginary part of $g^{-1}(x)$
\begin{equation}
\Im [g^{-1}(x+i\epsilon)] = \pi\gamma(x) = -\frac{\Im [g(x)]}{|g(x)|^2}  
=\frac{\pi\hat{\rho}(x)}{|g(x)|^2} \label{gm:cr}
\end{equation}
From equation \ref{cr:den}, we see that  $\pi s^2\sigma(s)$ is the 
imaginary part of $G\hspace{-3mm}/~^{-1}_M$. 
\begin{equation}
\pi s^2\sigma(s) = \Im [g\hspace{-2mm}/^{-1}(s)]=Z\Im[g^{-1}(s)]=Z\pi\gamma(s).
\end{equation}
Hence
\begin{equation}
\sigma(s) = \frac{Z\gamma(s)}{s^2} \equiv \frac{\gamma\hspace{-2mm}/(s)}{s^2} 
\label{cr:gamm}
\end{equation}

To calculate the renormalization constant $Z$, we  use $g(z)g^{-1}(z)=1$, 
letting the $z\rightarrow x_p$,
\begin{equation}
\frac1Z = 1 +  \int dx' 
\frac{\gamma(x')}{(x'-x_p)^2} + \cdots .\label{Z:inv} 
\end{equation}
Making the analytic continuation to Euclidean space, we have the Euclidean 
inverse two point Green function
\begin{equation}
\frac1{G(p)} = (p^2 +m^2) \left[ 1 
+ \int d\kappa^2 \frac{\gamma(\kappa^2)}{(\kappa^2 - m^2)(\kappa^2 + p^2)} 
\right]
\end{equation}
Expand in power of $p^2$, 
\begin{equation}
\frac1{(p^2 + \xi^2)} = \frac1{\xi^2}\left( 1-\frac{p^2}{\xi^2} +\frac{p^4}{\xi^4} + \cdots \right),
\end{equation}
we have
\begin{eqnarray}
G^{-1}(p^2) & = & m^2 \left( 1 +  
\int d\kappa^2 \frac{\gamma(\kappa^2)}{(\kappa^2 - m^2)\kappa^2} \right)\nonumber\\
& + & p^2 \left( 1 +  
\int d\kappa^2 \frac{\gamma(\kappa^2)}{\kappa^4} \right)\nonumber\\
& + & p^4 \left( -   
\int d\kappa^2 \frac{\gamma(\kappa^2)}{\kappa^6} \right) + 
\cdots \label{Gi:poly}\\
& \equiv & \frac1{Z'} \left(m'^2 + p^2 + (\cdots)p^4 + \cdots \right) \label{tp:inv}
\end{eqnarray}

In the last step, we define the $m'$ and $Z'$. These quantities can be 
measured on the lattice. By taking derivatives of equations \ref{tp:fn} and
\ref{tp:inv}, we can find relations of $m, m'$ and $Z, Z'$ as
\begin{equation}
m'^2 \geq m^2 ~,~~~~~~ Z' \geq Z  \label{Zm:ieq}
\end{equation}

\section{Upper Bound of Inclusive Cross Section}

From equation \ref{Gi:poly} and \ref{tp:inv}, one can calculate $\frac1{Z'}$

\begin{equation}
\left. \frac{\partial G^{-1}(p)}{\partial p^2}\right|_{p^2=0} = 
\frac1{Z'} \label{eq:dG1Z}
\end{equation}
So we have
\begin{equation}
\frac{1}{Z'}  =   1  + \int d\kappa^2 
~\frac{\gamma(\kappa)}{\kappa^4}  \label{Zp:inv}
\end{equation}

In equation \ref{cr:gamm}, we have defined the renormalized $\gamma(s)$ as
\begin{equation}
\gamma\hspace{-2mm}/ (s)= Z \gamma(s) = s^2\sigma(s). \label{eq:Zgamma}
\end{equation}
From the above equation  and \ref{Zp:inv}, one can derive an upper 
bound on the integrated inclusive cross section
\begin{eqnarray}
\int ds ~\sigma(s) & = & Z\int ds \frac{\gamma(s)}{s^2}  \nonumber \\
& = &  Z\left( \frac1{Z'} - 1 \right) \label{bd:Z}\\
& \leq &  \left( \frac1{Z'} - 1 \right) \label{bd:Zp}
\end{eqnarray}

Thus $1/Z' -1$, obtained from the lattice MC simulation of Euclidean $G(p^2)$, 
provides a rigorous upper bound on the inclusive cross section $\sigma(s)$. If
$1/Z' -1 $, computed non-perturbatively, is close to its weak coupling 
perturbative value, positivity implies that all of the multiparticle 
production cross sections are small. If $1/Z' -1 $ is large, then the
multiparticle production cross sections are making a large contribution.
However, the $1/Z' -1 $ only bounds the integral. Thus even if it is large, one
can not conclude that the production cross section for any particular $n$ 
is large. Note that from equation \ref{bd:Z} and \ref{Zm:ieq}, one can also
have
\begin{equation}
\int ds ~\sigma(s) \leq \frac{Z}{Z'} \leq 1
\end{equation}
This implies that if the theory is strongly coupled($Z<<1$), the inequality
in \ref{bd:Z} is an inefficient bound. For weak coupling case
($Z' \sim Z \sim 1$), the inequality in \ref{bd:Z} is nearly saturated.

\chapter{Critical Energy and the Relation to Euclidean Momentum}

  In this chapter, I will estimate the critical energy for which the 
multiparticle production turns on exponentially. First, I will calculate 
the multiparticle cross section using a statistical calculation of the 
multiparticle phase space integral and estimate the critical
energy from the amplitude. Then I will compare the result
with the direct phase space calculation for low order $n$. At the last, 
I will discuss
the relation of the critical energy with the momentum on the lattice and
determine the condition for the lattice simulation to be sensitive to the
hypothetical large cross section for high multiplicity events. 

\section{Amplitude for $f\bar{f} \rightarrow n\phi$}

In this section, we derive for the cross section for the production of $n$
scalar in $f\bar{f}$ annihilation and estimate the critical energy which leads
to sudden exponential growth.  Several authors\cite{Go:phyletb246,
Co:phyletb243,Vo:phyletb293,AKG:nphb341,Br:prd46} 
have attempted to estimate the cross 
section perturbatively for the production of large number of $\phi$. 
The most straightforward is the direct counting of graphs by a recursive
construction, which we will outline here for the case of the pure $\phi^4$
without the spontaneous symmetry breaking(SSB), in which the recursion 
relation is somewhat simpler. The calculation for the case with spontaneous 
symmetry breaking will be discussed later. 
For the one with SSB, we will follow Argyres, 
Kleiss, and Papadopoulo(AKP) and Voloshin's method
\cite{AKG:nphb341,Vo:phyletb293} and do the 
calculation in 1 + 2 dimensions.

\begin{figure}[tbp]
\begin{center}
\leavevmode
\epsfxsize 5in
\epsfbox{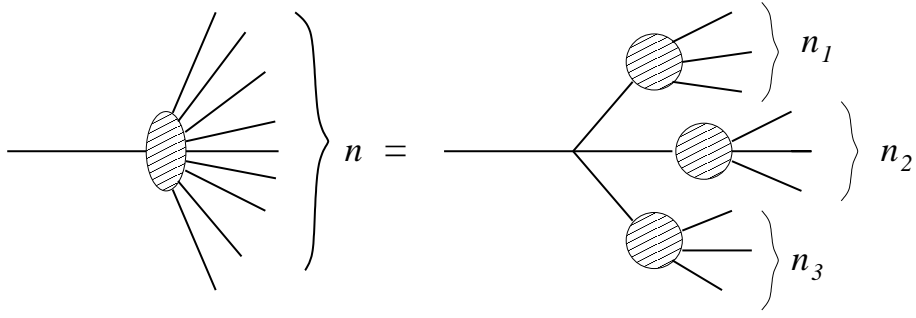}
\caption{Diagrammatic structure of the recursion relation for $\phi^4$ 
interactions in symmetry case} \label{figure:treesym}
\end{center}
\end{figure}

Consider $f\bar{f}$ annihilation into a
single off-shell $\phi$, which subsequently decays into $n$ on-shell $\phi$.
Throughout the calculation, we only consider tree-level perturbation theory
and neglect
the contributions from $f\bar{f}$ annihilation into more than just one off-shell $\phi$.
In the pure $\phi^4$ theory, symmetries phase, we can denote the 
number of topologically distinct diagrams by $a(n)$.(see Figure 
\ref{figure:treesym})
\begin{equation}
a(n) = \lambda\frac{n!}{3!} \sum_{n_1,n_2,n_3,odd} \delta_{n_1+n_2+n_3,n}
D(n_1)D(n_2)D(n_3)\frac{a(n_1)}{n_1!}\frac{a(n_2)}{n_2!}\frac{a(n_3)}{n_3!}
\end{equation}
The $n!/(n_1!n_2!n_3!)$ counts the number of ways in which the $n$ final 
particles can be grouped in to three sets of $n_1,n_2$ and $n_3$ particles.
the $\delta$ symbol accounts for that fact that $n_1+n_2+n_3=n$ and the
factor $D(n_1)D(n_2)D(n_3)$ is given by the propagators $D$ connecting the 
root vertex with the subtrees with $n_1,n_2$ and $n_3$ particles in each.
Note that each $a(n_i)$ can also subdecay into another three $a(n_j)$ 
recursively. 

   We consider the 
possible momentum configurations for the amplitude. We can
find upper and lower bounds at the extreme 
cases of the allowed momentum configurations. For a generic momentum 
configuration, the recursion relation for the amplitude $a(n)$ is too
complicated too deal with. We make a simplifying assumption that leads to a 
tractable problem. We assume that for each partition of $n$ momentum  into
$n_1+n_2+n_3$ momentum, the average $\langle p_i\cdot p_j \rangle$ is the 
same for each of the momentum configuragions associated with $n_1$, $n_2$,
and $n_3$. The propagator for each internal line is the same.

\begin{eqnarray}
D^{-1}(n)=q^2-m^2 & = & (p_1 + p_2 + \cdots + p_n)^2 -m^2 \nonumber \\
& = & n p_i^2 + {\sum_{i,j}}' p_i\cdot p_j - m^2 ~~~~~(i \neq j) \nonumber \\
&=& n(n-1) \langle p_i \cdot p_j\rangle + (n-1) m^2 \nonumber \\
& = & n^2 \langle p_i \cdot p_j \rangle - n(\langle p_i \cdot p_j 
\rangle-m^2) - m^2 \\ 
& \simeq & (n^2 -1) \langle p_i \cdot p_j\rangle  \label{kn:eqp}
\end{eqnarray}
We write
\begin{equation}
\langle p_i\cdot p_j \rangle = \xi^2 m^2  \label{p2:dot}
\end{equation}
where $\xi^2$ is just a positive constant. Clearly if $\xi^2 = 1$, all
3-momenta $\vec{p}_i$ must be zero and the phase space volume vanishes. This 
condition gives us a trivial lower bound on the cross section. 
For $\xi^2 > 1$, we have
\begin{equation}
D(n_i) = \frac{1}{(p_1+p_2+\cdots+p_i)^2 -m^2} = \frac1{(i^2-1)\xi^2 m^2}
\end{equation}
and the recursion relation for the amplitude becomes
\begin{eqnarray}
a(n) & = & \frac{\lambda}{3!} \sum_{n_1,n_2,n_3,odd} \delta_{n_1+n_2+n_3,n}
D(n_1)D(n_2)D(n_3)\frac{n!}{n_1!n_2!n_3!}\times \nonumber \\
& & \frac{a(n_1)a(n_2)a(n_3)}
{(n_1^2-1)\xi^2 m^2(n_2^2-1)\xi^2 m^2(n_3^2-1)\xi^2 m^2}
\end{eqnarray}
The solution of the above recursion equation for $a(n)$ is
\begin{equation}
a(n) = - \xi^2 m^2 (n^2-1) n! \left(\frac{\lambda_4}{48\xi^2 m^2} 
\right)^{\frac{n-1}{2}} \label{am:symm}
\end{equation}
Note that, we have approximated the multiparticle 
amplitude depending on $p_1,p_2\cdots p_n$ to an amplitude which is a 
function of total energy(or total number of out going particles) only.

For the broken symmetry case, the interaction term in Lagrange is
\begin{equation}
{\cal L}_I = \frac{\lambda_3}{3!}\phi^3 + \frac{\lambda_4}{4!}\phi^4
\end{equation}
with
\begin{equation}
\lambda_3 = m \sqrt{3\lambda_4}~,~~~~~~~~\phi = v + \hat{\phi}
\end{equation}
\begin{figure}[tbp]
\begin{center}
\leavevmode
\epsfxsize 6in
\epsfbox{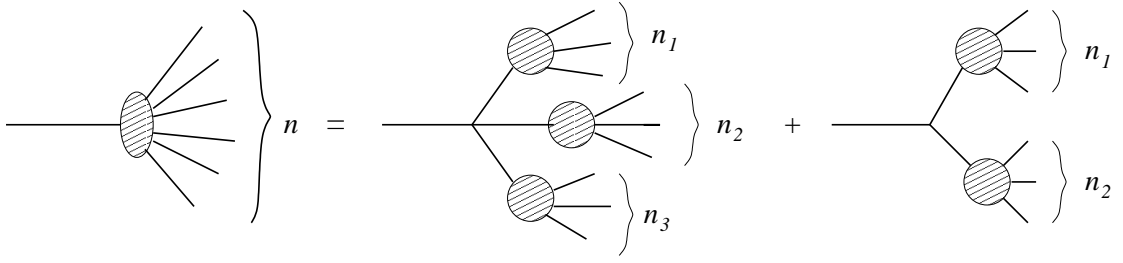}
\caption{Diagrammatic structure of the recursion relation for mixed $\phi^4$
and $\phi^3$ interaction} \label{fig:treebs}
\end{center}
\end{figure}
Use the same kinematic assumption in equation \ref{kn:eqp} and \ref{p2:dot}, 
the recursion relation for the amplitude becomes
\begin{eqnarray}
a(n) & = & -i\lambda_4 \frac{n!}{3! m^6\xi^6} \sum_{n_1,n_2,n_3}
\delta_{n_1+n_2+n_3,n}
\frac{ia(n_1)}{n_1!(n_1^2-1)}\frac{ia(n_2)}{n_2!(n_2^2-1)}
\frac{ia(n_3)}{n_3!(n_3^2-1)} \nonumber \\
&& -i\lambda_3 \frac{n!}{2! m^4\xi^4} \sum_{n_1,n_2} \delta_{n_1+n_2,n}
\frac{ia(n_1)}{n_1!(n_1^2-1)}\frac{ia(n_2)}{n_2!(n_2^2-1)}
\end{eqnarray}

We can make an ansatz \cite{AKG:nphb341}:
\begin{equation}
a(n) = -im^2\xi^2n!(n^2-1)b(n)
\left(\frac{\lambda_4}{m^2\xi^2}\right)^{\frac{(n-1)}{2}}, \label{am:ssb}
\end{equation}
Which leads to the following equation for the $b(n)$:
\begin{equation}
b(n) = \frac{1}{(12)^{(n-1)/2}}\left( (\frac{\mu}{\xi} + \sqrt{3})^n - 
(\frac{\mu}{\xi} - \sqrt{3})^n \right)
\end{equation}
with $b(1)=1$. The $\mu$ in the case of spontaneous symmetry breaking is
\begin{equation}
\mu = \frac{\lambda_3}{m\sqrt{\lambda_4}} = \sqrt{3}
\end{equation}

In the special case $\lambda_3 = 0$, one can see that
the amplitude \ref{am:ssb} goes back to equation \ref{am:symm}.
Once again, the amplitude of multiparticle production has been approximated 
by an amplitude which is a function of total energy only.

\section{The Phase Space}

In chapter 3, we have defined the transition probability as
\begin{equation}
\Gamma(n,E) = \frac{1}{m n!} \int d\Phi_n |A(p_1,\cdots,p_n)|^2
\end{equation}
In last section, we have obtained an
approximation of the $1 \rightarrow n$ amplitude.  
$$
A(p_1,p_2,\cdots,p_n) \longrightarrow A(E)
$$
and the problem has been simplified to the amplitude $A(E)$ times the 
volume of the phase space.
In order to calculate the transition probability and the cross section, 
we have to calculate the volume of the 
multiparticle phase space in as accurate a manner as possible. When
$n$ becomes very large, the exact calculation becomes very difficult. 
So, we still
need to find a good approximation method to calculate the phase space
integral. One
method is to use a statistical method which was developed by F.
Lurcat and P. Mazur \cite{Nu:ci64} and Fermi \cite{PTP:1950}. We will derive 
the phase-space volume based on their method. To simplify the calculation, we
choose the center of mass frame(CM) for our phase space integral.
$$
P^\mu = (P^0,\vec{P}) = (E,0)
$$
where
$$ P^\mu = \sum^n_i p^{\mu}_i$$

In general, the phase space integral in three dimensions has the following 
form:
\begin{equation}
\int d\Phi_n  =  \frac1{n!}\int \frac{d^3p_1}{(2\pi)^3}(2\pi)\delta(p^2_1-m^2)
\cdots 
\int \frac{d^3p_n}{(2\pi)^3}(2\pi)\delta(p^2_n-m^2) 
(2\pi)^3\delta^3(\vec{P}-\sum p_i)  
\end{equation}
We rewrite the phase space integral as
\begin{eqnarray}
\int d\Phi_n & = & \frac1{n!} \int d(PS)_n(E) \\
(PS)_n(E) & = & \frac{R_n(E)}{(2\pi)^{2n-3}}  \label{PS:Rn}
\end{eqnarray}
and 
\begin{equation}
R_n(p) = \prod^n_{r=1} \int d^3p_r \delta(p^2_r - m^2)\theta(p^0_r-m)
\delta^3(p-\sum_r p_r)
\end{equation}
We use a Laplace transformation to factorize this
\begin{eqnarray}
{\cal R}_n(a) & = & \int d^3p e^{-a\cdot p} R_n(p) \nonumber \\
&=& \prod^n_{r=1} \int d^3p_r e^{-a\cdot p_r}\delta(p^2_r-m^2)\theta(p^0_r-m)
\nonumber \\
&\equiv & \left( m \varphi_0(a) \right)^n
\end{eqnarray}
where 
\begin{equation}
m\varphi_0(a) = \int d^3p_i~e^{-a\cdot p_i}\delta(p^2_i-m^2)\theta(p^0_i-m)
\label{vr:phi0}
\end{equation}
for each single particle.
We can define the single particle probability distribution function({\bf spd}) 
of momentum in phase space as
\begin{equation}
u(a,p_i) = \frac{e^{-a\cdot p_i}\delta(p^2_i-m^2)\theta(p^0_i-m)}{m\varphi_0(a)}
\end{equation}
so that 
\begin{equation}
\int d^3p_i~u(a,p_i) = 1
\end{equation}
Suppose the $n$ individual momentum $p_{1,2\cdots,n}$ are independent 
variables and $P=p_1+p_2+\cdots+p_n$. From 
the statistical theory, we can define the full(all particles) distribution 
function as
\begin{eqnarray}
U_n(a,P) & = & \frac{R_n(P)e^{-a\cdot P}}{{\cal R}_n(P)} 
= \frac{R_n(P)e^{-a\cdot P}}{(m \varphi_0(a))^n} \label{Un:Rn}\\
&=& \prod^n_{r=1} \frac{\left( \int d^3p_r\delta(p^2_r-m^2)\theta(p^0_r-m)
e^{-a\cdot p} \right) \delta^3(P^{\mu} - \sum_r p^{\mu}_r)}
{(m\varphi_0(a))^n} \\
&=& \prod^n_{r=1} \left( \int d^3p_r~u(a,p_r) \right) 
\delta^3(P^{\mu} - \sum_r p^{\mu}_r)
\end{eqnarray}
We have
\begin{equation}
\int d^3p~U_n(a,P) = 1, ~~~~~~~~U_n(a,P)=\prod^n_r~u(a,p_r)
\end{equation}
From equation \ref{PS:Rn} and \ref{Un:Rn}, we see that if we know
the distribution function $U_n(a,p)$, we can calculate the $R_n(a,p)$ and
the phase space $(PS)$ integral.  The $U_n(a,p)$ can be calculated by using
the central limit theorem.  For large $n$, the sum of the $n$ 
independent variables($P=p_1+p_2+\cdots+p_n$) will have a probability
distribution $U_n(a,P)$ approaching to the 
Gaussian. So we can expand $U_n(a,P)$ as
\begin{equation}
U_n(a,P)=\Psi_G(P)\left( 1 + \frac{C_1}{\sqrt{n}} + \frac{C_2}{n} + \cdots
\right)
\end{equation}
where
\begin{equation}
\Psi_G(P)=\prod_{\mu=0,1,2} \frac1{\sqrt{2\pi (\bigtriangleup P)}}~
e^{\left(-\frac{(p^{\mu}-\langle P^{\mu}\rangle)^2}{2\bigtriangleup P^\mu}\right)}
\end{equation}
The $C_1$, $C_2$ are constants and $\bigtriangleup P^\mu$ is the variation 
of $P^\mu$, and $\langle P^{\mu}\rangle$ is the expectation value of 
$P^\mu$. In the CM
frame, $P^\mu=(P^0,0)$ and $a\cdot P = (aP^0,0)$. The trick is now to obtain a
closed-form approximation for $\Psi(P)$ and then to put $P^\mu$ to its
desired value $P^0=\sqrt{s}$, $\vec{P}=0$. The expectation value
of $P^1$ and $P^2$ is zero($\langle P^1\rangle =\langle P^2\rangle =0$). Then
\begin{eqnarray}
\langle p^0\rangle & = & nm\frac{\varphi_1(a)}{\varphi_0(a)} \nonumber \\
\bigtriangleup P^0 &=& nm^2\left[\frac{\varphi_2(a)}{\varphi_0(a)} - 
\left( \frac{\varphi_1(a)}{\varphi_0(a)} \right) ^2 \right] \nonumber \\
\bigtriangleup P_1 = \bigtriangleup P_2 & =& \frac{nm^2}{2} \left( 
\frac{\varphi_2(a)}{\varphi_0(a)} -1 \right) 
\end{eqnarray}
with the definition of $\varphi_1$, $\varphi_2$ as
\begin{eqnarray}
m^2\varphi_1(a) & = & \int d^3p_i~\delta(p^2_i-m^2)\theta(p^0_i-m)e^{-ap^0_i}
p_i^0 \label{vr:phi1}\\
m^3\varphi_2(a) & = & \int d^3p_i~\delta(p^2_i-m^2)\theta(p^0_i-m)e^{-ap^0_i}
(p_i^0)^2 \label{vr:phi2}
\end{eqnarray}
In the CM frame, $P^\mu=(P^0,0)$, the expectation value of $P^1,P^2$ is zero. 
Note that the asymptotic series of $U_n(a,P)$ is formally good for any 
$P^\mu$ but useful only very close to the peak of the Gaussian. We therefore
can tune the constant $a$ such that $\langle P^0\rangle=\sqrt{s}$. So,
if we choose $\langle P^0\rangle =\sqrt{s}=P^0$, the exponential function 
equal one and the multiparticle distribution function becomes
\begin{equation}
U_n(a,P)=\frac2{(2\pi)^{3/2} (nm^2)^{3/2}
\left( \frac{\varphi_2(a)}{\varphi_0(a)} -1 \right) 
\left[\frac{\varphi_2(a)}{\varphi_0(a)} -
\left( \frac{\varphi_1(a)}{\varphi_0(a)} \right) ^2 \right]^{1/2} }
\end{equation}
and from equation \ref{Un:Rn}, we have
\begin{equation}
R_n(P)=\frac{2(m\varphi_0(a))^n~e^{aE}}
{(2\pi)^{3/2} (nm^2)^{3/2} 
\left(\frac{\varphi_2(a)}{\varphi_0(a)} -1 \right)
\sqrt{\frac{\varphi_2(a)}{\varphi_0(a)} -
\left( \frac{\varphi_1(a)}{\varphi_0(a)} \right) ^2} } \label{eq:Rn}
\end{equation}
So the phase space integral becomes
\begin{equation}
(PS)_n(E)= \frac{2(nm^2)^{-3/2}(m\varphi_0(a))^n~e^{aE}}
{(2\pi)^{2n-3/2}\left(\frac{\varphi_2(a)}{\varphi_0(a)} -1 \right)
\sqrt{\frac{\varphi_2(a)}{\varphi_0(a)} -
\left( \frac{\varphi_1(a)}{\varphi_0(a)} \right) ^2}} \label{PS:PS}
\end{equation}
The $\varphi_0$, $\varphi_1$, and $\varphi_2$ can be calculated directly from
equation \ref{vr:phi0},\ref{vr:phi1},and \ref{vr:phi2}
\begin{eqnarray}
\varphi_0(a) & = & \frac{\pi~e^{-am}}{am} \\
m^2\varphi_1(a) & = & \left( \frac1{a} + m \right) m\varphi_0 \\
m^3\varphi_2(a) & = & \left[ \frac1{a^2} + (\frac1{a} + m )^2 \right] m\varphi_0
\end{eqnarray}

Note that we get the equation \ref{eq:Rn} and \ref{PS:PS} by using a statistical
method which is accurate for large $n$. Although our interest is in large $n$, 
is interesting to compare the $R_n$ with the exact phase space integral at 
$n=3$. For convenience, we define $1/ \nu$ as the ratio of $E$ and $nm$
\begin{equation}
\frac{E}{nm} = \frac1\nu = \frac{\varphi_1}{\varphi_0} \label{Emnu}
\end{equation}
The statistical method gives us
\begin{eqnarray}
R'_3 &=& \left(\frac{\pi}{6}\right)^{3/2}(1-\nu)~e^3  \\
& \sim & 7.61 \times (1-\nu) \nonumber
\end{eqnarray}
and the exact result is
\begin{eqnarray}
R_3 & = & \frac{\pi^2}2 (1-\nu)  \\
& \sim & 4.93 (1-\nu) \nonumber
\end{eqnarray}
One can see that even at small $n$, the form of the dependence on $\nu$ is the
same adn the coeficients are the same order of magnitude. For slightly larger 
$n$, the results converge rapidly.

\section{The Critical Energy and The relation to Lattice Momentum}

The critical energy is the energy at which the transition rate begins to grow 
exponentially.
In section 4.1 and 4.2, we have obtained the ingrediets to approximate the 
transition rate as an amplitude squared times the phase space volume in 
CM frame. 
\begin{equation}
m\Gamma(n,E) = |a(n)|^2 \frac1{n!} \int d\Phi_n
\end{equation} 
From equation \ref{am:ssb} and \ref{PS:PS}, the $n$ particle amplitude for
the broken symmetry case is
\begin{equation}
A(n) \simeq -\xi^2m^2(n^2-1)n!\hat{b}(n)\left(\frac{\lambda_4}{48\xi^2m^2}
\right)^{\frac{n-1}{2}}
\end{equation}
and
\begin{equation}
\int d\Phi_n(E) \simeq \frac{2(nm^2)^{-3/2}(m\varphi_0(a))^n~e^{aE}}
{(2\pi)^{2n-3/2}\left(\frac{\varphi_2(a)}{\varphi_0(a)} -1 \right)
\sqrt{\frac{\varphi_2(a)}{\varphi_0(a)} -
\left( \frac{\varphi_1(a)}{\varphi_0(a)} \right) ^2}}
\end{equation}
where in broken symmetry case, $\lambda_4=6\lambda$ and $\lambda_3=\sqrt{3}$. 
So
\begin{equation}
\hat{b}(n) = \frac12 \left[ (1+\frac1{\xi})^n - (-1 + \frac1{\xi})^n \right]
\end{equation}
Again,  we choose the CM frame condition $\langle P^0\rangle =P^0=\sqrt{s}=E$. 
With $\nu$ defined in \ref{Emnu}
and keeping just the exponential behavior, 
one can find the $n$ particle transition rate is
\begin{eqnarray}
\Gamma(n,E) & \simeq &  \frac{\pi m^2}{2e}\left[ (1-\nu)^2\times ng \nu(1-\nu)
\right]^n \\
& = & \frac{\pi m^2}{2e} e^ {\left[ n(2\ln(1+\nu) + \ln(ng) + \ln(\nu) +
\ln(1-\nu)) \right]}
\end{eqnarray}
where  we define the coupling constant
\begin{equation}
g = \frac{\lambda}{32\pi m}
\end{equation}
and from  $E^2\simeq n^2\xi^2 m^2$, we have
\begin{equation}
\xi = \frac1{\nu}.
\end{equation}
One can see that the tree graph calculation leads to an exponentially abrupt
rise of the inclusive cross section at some critical energy. If we fix $n$ but
send $E$ to the infinity, $\nu$ will vanish as well as the transition
rate. If we fix $E$, then $n$ can not goes to infinity because there is not
enough energy. To find the critical energy for both $E$ and $n$ very
large, we can keep the ratio $\nu$ fixed at its maximum value and find the
minimum energy of the threshold. The threshold for
this blow up happens when the exponent is positive. So the critical energy 
$E_*$ is the root of the exponent. Set the energy parameter
$\epsilon = gE/m$ so that 
\begin{equation}
\epsilon\nu = ng
\end{equation}
Exponential growth occurs for
\begin{equation}
\epsilon \nu^2(1-\nu)(1+\nu)^2 \geq 1.
\end{equation}
The function $\nu^2(1-\nu)(1+\nu)^2$ has maximum value 0.431, so that 
the minimum value for $\epsilon$ is $\epsilon_*=2.3198$.  The
critical energy is
\begin{equation}
E_*=2.3198 \frac{m}{g} \label{eq:Ecrit}
\end{equation}

In 1 + 2 dimension, the spectral integral \ref{eq:spctrl} is convergent, so the 
dominant contributions to the spectral representation of $G(p^2)$,(Equation
\ref{tp:fn}), come from  $\hat{\rho}(\kappa^2)$ with $\kappa^2 \leq p^2$. 
When we compute $G(p^2)$ on the lattice, the Euclidean lattice momenta are
restricted to the Brillouin zone.
\begin{equation}
-\frac{\pi}{a} \leq p_i \leq \frac{\pi}{a}
\end{equation}
The restriction for $p^2$ will be
\begin{equation}
0 \leq p^2 \leq D\left( \frac{\pi}{a} \right)^2
\end{equation}
where $D$ is the dimension of the lattice and $a$ is the lattice constant.
Thus for the lattice calculation to be sensitive to $\sigma(s)$ for $s$ up
to $E^2_*$, we must work on a lattice satisfying
\begin{equation}
E^2_* \leq D\left( \frac{\pi}{a} \right)^2
\end{equation}
Dividing by $m^2$, the requirement is 
\begin{equation}
D\frac{\pi^2}{m^2_L} > \left( \frac{E_*}{m} \right)^2.
\end{equation}
where $m_L = ma$ is the mass in lattice units, which is measurable on the
lattice, and tends to zero as the bare parameters are tunes to approach 
criticality(continuum limit). Define the correlation length in the lattice
units $\xi=1/m_L$ which should satisfy
\begin{equation}
\frac{E_*}{\sqrt{3}\pi m} \leq \frac1{m_L} = \xi_L \label{eq:Eml}
\end{equation}
in three dimensions.

For  a finite lattice calculation to give a good approximation to the 
continuum results in general, we should have
\begin{equation}
a \ll \xi = \frac1{m} \ll L=Na
\end{equation}
Use equation \ref{eq:Ecrit},\ref{eq:Eml} one can find the condition
\begin{equation}
1 \ll 74.24\frac{m}{\sqrt{3}\lambda} < \xi \ll N \label{xi:cond}
\end{equation}

This set of conditions restricts the range of parameters of the simulation. 
In order to satisfy the critical energy threshold, the correlation length 
from the simulation should longer than $\sim 42.86\frac{\lambda}{m'}$. 
For the $\lambda /m' \sim 1$ in the broken symmetry phase, 
this set of conditions require lattices of size $128^3$ or $256^3$.

\chapter{Lattice Simulation}

Monte Carlo simulations have become increasingly important as a tool
in studying complex statistical systems and fundamental properties
of quantum field theory. 
In this chapter, we will discuss how to measure $Z'$ and $m'^2$ on the 
lattice  using Monte Carlo method. Because we are working close to
the critical line, we experience the critical slowing down problem. I will
discuss the cluster update algorithm in section two. We also want to 
know the properties of our simulation when the lattice spacing goes to zero. 
So I will discuss the continuum limit of the simulation in section three.
In section four, we will specify the simulation method.  

\section{Monte Carlo Simulation}

The Monte Carlo simulation on the lattice is well known. We will just follow
the standard way\cite{Lt:QFT}. However, it is useful to review the major 
features of the three-dimensional theory in the parameter space we use. 
The Euclidean action is discretized on an $N^3$ lattice with lattice 
constant $a$ and periodic boundary conditions. The continuum mass and coupling
$\lambda$ are related to the lattice parameters by $\lambda_L=\lambda a$
and $m_L=ma$. The discretized action is 
\begin{equation}
S=\sum_{\vec{n}} \left\{ \frac12\sum_{\hat{n}}
\left[ \phi(\vec{n}+{\bf \hat{n}}) - \phi(\vec{n}) \right]^2
+\frac12\mu^2_L\phi(\vec{n})^2+\frac14\lambda_L\phi(\vec{n})^4 \right\} ,
\end{equation}
where $\mu^2_L$ is the lattice bare mass squared. The discretized Lagrangian
can be written as
\begin{eqnarray}
{\cal L} & = & -\frac12 \sum_{\vec{n}}\left[\phi(\vec{n}+\hat{n})\phi(\vec{n}) 
+ \phi(\vec{n})\phi(\vec{n}-\hat{n}) \right] \nonumber \\
& & + \sum_{\vec{n}} (3+\frac{\mu^2_L}{2}) \phi(\vec{n})^2 
+ \frac{\lambda_L}{4} \sum_{\vec{n}} \phi(\vec{n})^4 
\label{disLarg}
\end{eqnarray}
Since the momentum on the lattice should be inside the Brillouin zone, 
the discretized free propagator is
\begin{equation}
G_0(k) = \frac1{\left\{ \sum_{i=1}^3 \frac1{a^2} 2\left( 1 - \cos(ak_i) \right) 
+ m^2 \right\} } \label{G0:latt}
\end{equation} 
Note that in the simulation, we didn't distinguish different lattice
momentum vectors which give the same momentum magnitude, {\em i.e.}, 
$|k_{i,j,k}|=|k_{j,k,i}|=|k_{k,i,j}|$... etc.

\begin{figure}[tbp]
\begin{center}
\leavevmode
\epsfxsize 5in
\epsfbox{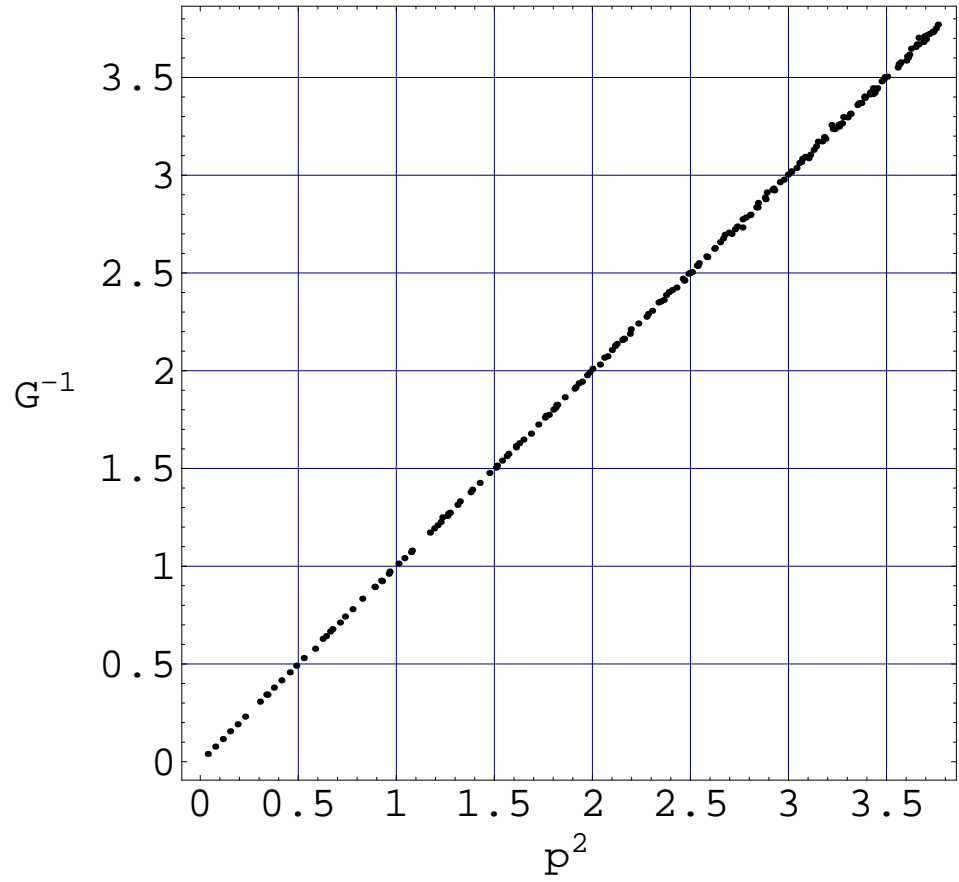}
\caption{Result of Monte Carlo evaluation of $G^{-1}_0(\vec{p})$ plotted
against exact analytic massless 
$G^{-1}_0(\vec{p}) = 4\sum_\mu \sin^2(p_{\mu})^{-1} \equiv 
\vec{p}^{~2}$ at $32^3$ lattice. Error bars on Monte Carlo results are too
small to show up on this plot.}
\label{figure:freeprop}
\end{center}
\end{figure}

We use the Metropolis algorithm to generate a sequence of equilibrium
configurations. Here we only consider the nearest neighbor interaction
at each lattice site. The probability density is
\begin{equation}
dP(\phi) = \frac{e^{-S(\phi)}}{\cal Z} \prod_{\vec{n}}d\phi(\vec{n})
\end{equation}
where the partition function is 
\begin{equation}
{\cal Z} = \prod_{\vec{n}}\int d\phi(\vec{n})~e^{-S(\phi)}.
\end{equation}
A random walk goes through the phase space with the acceptance probability
\begin{equation}
e^{i(S_{new} - S_{old})}
\end{equation}
for each change of $\phi \rightarrow \phi + d\phi$ at each lattice site.
We measure the one-point and two-point functions after sweeping through the 
lattice.
We also compute the Fourier transform of the $\phi$ field and calculate the
two-point function in momentum space by
\begin{equation}
G(k) = \langle \phi(k)\phi(-k) \rangle .
\end{equation}
Then at small momentum $k$, the reciprocal of the two point function is
\begin{equation}
G(k)^{-1} = \frac1{Z'}(m^2+k^2+O(k^4)\cdots)
\end{equation}
We then plot the reciprocal of the two-point function in momentum space
against the reciprocal of the Fourier transform of the massless free lattice
two-point function ($m=0$ in \ref{G0:latt}). The quantities $m^2_L$ and
$Z'$ in equation \ref{tp:inv} are extracted from the intercept and slope at the
origin of this plot. To illustrate the plot, we have shown one example
in Figure \ref{figure:freeprop} . Here we plot the Monte Carlo evaluation of 
the free propagator($\lambda = 0$) $G^{-1}(\vec{p})$ against the 
massless $G^{-1}_0(\vec{p}) = 4\sum_\mu 
\sin^2(p_{\mu})^{-1} \equiv \vec{p}^{~2}$. One can see that the Monte Carlo
results is the same as the known analytic result equation \ref{G0:latt}.
Note that this provids a nontrivial check of a large fraction of our
Fortran programs.  

\section{Cluster Update Algorithm}

The Metropolis algorithm has a major drawback when working very close to 
the critical point. It suffers from critical slowing down if one approaches
the phase transition. As one approaches to the critical point, both the 
correlation length and correlation time diverge. The relation of the
correlation length $\xi$ and correlation time $\tau$ are\cite{HH:dynz,Lt:QFT} 
\begin{equation}
\tau \sim \xi^z \sim L^D
\end{equation}
where the $z$ is so called dynamic exponent and the $\tau$ is measured in 
computer steps. This equation tells us how the time correlation gets longer 
as the correlation length diverges near criticality. However, in a system 
of finite size, the correlation length can never really diverge. With the 
lattice size $L$, the correlation time increase as
\begin{equation}
\tau \sim L^z
\end{equation}
when close to the criticality. One can see that as we approach 
criticality, we need to have longer CPU time to get the required statistics.

To avoid the critical slowing down, one must define appropriate nonlocal(or
collective) variables and a new dynamics for driving them. Progress has 
been made with some nonlocal algorithms for both discrete spin models and 
continuum fields. Swendsen and Wang\cite{SW:cls} have used the Fortuin-Kastelyn
\cite{FK:pha} percolation map for the Potts model to define collective
coordinates that allow domains to be inverted with zero free energy cost.
Another approach was proposed by Wolff\cite{Wf:prl23} which is 
essentially the same as
Swendsen and Wang algorithm with some modifications.
Further improvements are possible by introducing multigrid methods\cite{KD:mg}.

However, the algorithm for the $\phi^4$ field theory is a little more 
complicated. Brower and Tamayo\cite{BT:phi4cls}  proposed an algorithm for
$\phi^4$ field theory based on the Swendsen and Wang algorithm. Here we will 
follow Brower and Tamayo's method but switch to the Wolff algorithm. Table 
\ref{dynexp}
compare the dynamic exponent value for the Ising model in different dimension 
and algorithm. One can see that the Wolff algorithm is a little better
than Swendsen and Wang algorithm in three dimension.
\begin{table}[tbp]
\caption{Comparison of the values of the dynamic exponent $z$ for
different algorithms for Ising model in various dimensions\cite{dy:z}.} 
\label{dynexp}
\begin{center}
\begin{tabular}{|c|c|c|c|}
\hline
dimension d & Metropolis & Wolff & Swendsen-Wang \\
\hline
2 & 2.167 $\pm$ 0.001 & 0.25 $\pm$ 0.01 &  0.25 $\pm$ 0.01 \\
3 & 2.02 $\pm$ 0.02 & 0.33 $\pm$ 0.01 & 0.54 $\pm$ 0.02 \\
4 & - & 0.25 $\pm$ 0.01 & 0.86 $\pm$ 0.02 \\
\hline
\end{tabular}
\end{center}
\end{table}

In our simulation, the
update algorithm consists basically of two parts:{\bf (1)} a conventional
Metropolis Monte Carlo update for the $\phi(x)$ field and {\bf(2)} Wolff cluster
identification and flipping. To identify the cluster of $\phi$ field,
one can introduce a discrete variable $s_{\vec{n}}$ to
describe the Ising like property of the field $\phi$,
\begin{equation}
\phi(\vec{n}) = s_{\vec{n}} |\phi(\vec{n})|
\end{equation}
where the $s_{\vec{n}}=\pm 1$. 

The detailed procedure is as follows:
\begin{description}
\item[(1).] Update the $\phi(\vec{n})$ fields via a standard local Monte
Carlo algorithm. In our case, it is the Metropolis algorithm.
\item[(2).] Choose a site $\vec{n}$ at random from the lattice.
\item[(3).] At that chosen site, $\vec{n}$, introduce an effective
Ising-type system,
\begin{equation}
e^{-S_{Ising}} = \prod_{\vec{n},\vec{n}+\hat{n}}~
e^{|\phi(\vec{n})\phi(\vec{n}+\hat{n})|(s_{\vec{n}}s_{\vec{n}+\hat{n}}-1)} 
\label{Z:Ising}
\end{equation} 
and look in turn at each of the neighbor of that $\phi$. Add the neighbor 
$\phi(\vec{n}+\hat{n})$ to the cluster with the probability
\begin{eqnarray}
P_{add} & = & 1 - e^{-|\phi(\vec{n})\phi(\vec{n}+\hat{n})|(1+s_{\vec{n}}s_{\vec{n}+\hat{n}})} \\
& = & 1 - e^{-(|\phi(\vec{n})\phi(\vec{n}+\hat{n})| +
\phi(\vec{n})\phi(\vec{n}+\hat{n})) } \label{Padd}
\end{eqnarray}
Note that $P_{add}=0$ if $\phi_{\vec{n}}$ and $\phi_{\vec{n}+\hat{n}}$ have 
opposite sign. That is, we only add the neighbors to the cluster when they 
have the same sign as the chosen one. 

\item[(4).] From the boundary of the cluster, repeat the adding step in (3) as 
many times as necessary until there are no 
$\phi$ left in the cluster whose neighbors have not been considered for
inclusion in the cluster.

\item[(5).] Flip the sign of the cluster.

\end{description}

\begin{figure}[tbp]
\begin{center}
\leavevmode
\epsfxsize 4in
\epsfbox{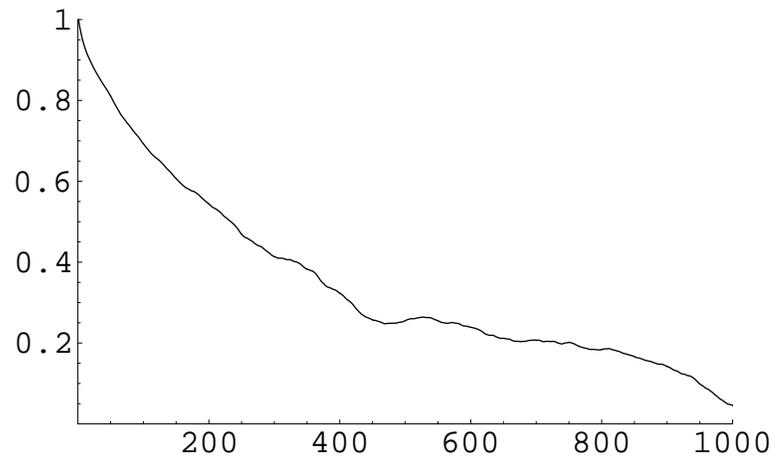}
\vskip 2cm
\epsfxsize 4in
\epsfbox{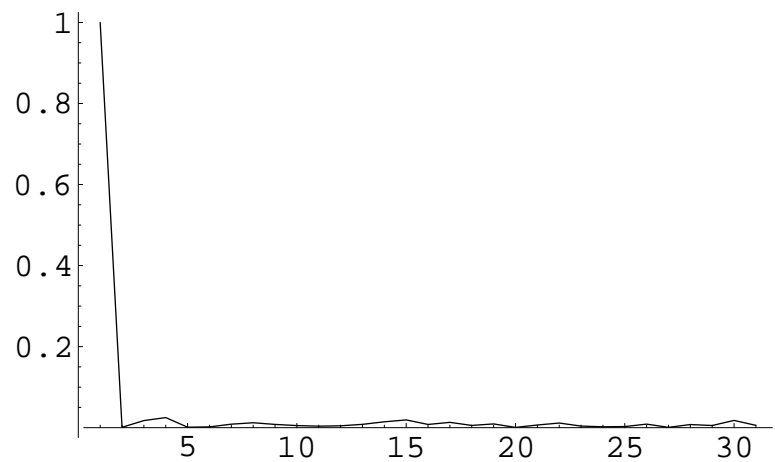}
\caption{The autocorrelation for the $\langle \phi \rangle$ with and 
without the cluster update algorithm. The $e^{-1}$ point for the 
one without the cluster update algorithm is at about 400(top).
The autocorrelation length for the one with cluster update algorithm is about
2 only(bottom).}
\label{fig:corr}
\end{center}
\end{figure}

The beauty of the Swendsen-Wang and Wolff update step is that the entropy
of the percolated cluster formation exactly cancels the surface energy to 
give zero free-energy cost. Detailed balance is guaranteed by the condition
that summing out the percolation variables in the joint distribution, 
$P_{add}$, of equation \ref{Padd} yields the correct Ising distribution of 
equation \ref{Z:Ising}. With the Wolff algorithm, our autocorrelation of
the VEV has been reduced from $\sim$400 to $\sim$2. Figure \ref{fig:corr}
shows the differences of the simulation results with
and without the cluster update algorithm. The simulation without the cluster
update algorithm has six millions sweeps whereas the one with the cluster 
update only needs hundred thousand sweeps.

\section{Continuum Limit}

The continuum $\phi^4$ quantum field theory can be recovered from a discrete
lattice theory by taking the limit of the lattice constant $a$ going to zero 
and possibly
rescaling the field strength by a factor of $Z(a)$. The existence of the 
limit of the lattice $\phi^4$ theory in three dimension has been proved by R. 
Fern\'{a}ndez, J. Fr\H{o}hlich and A. Sokal\cite{FF:rwqft}, and J, Glimm and 
A. Jaffe \cite{GJ:qphy}. To make use of  the simulation, we have to know the
behavior of the theory as $a \rightarrow 0$.

\begin{figure}[tbp]
\begin{center}
\leavevmode
\epsfxsize 6in
\epsfbox{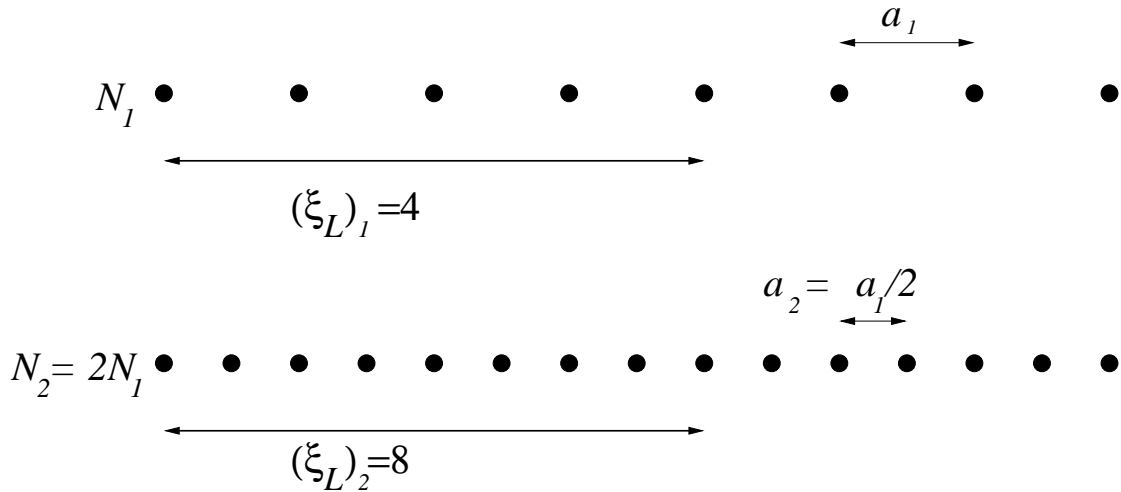}
\caption{The scaling of the continue limit. When we halve the lattice constant,
$a \rightarrow a/2$, the correlation length, $\xi_L$, will be doubled $\xi_L 
\rightarrow 2\xi_L$} 
\label{fig:scaling}
\end{center}
\end{figure}

In $\phi^4$ theory on the lattice, we have defined the $\lambda_L=\lambda a$
and $m_L=ma$ which measured in lattice units. The correlation length in 
physics units is $\xi=1/m$. The correlation length of the lattice units is
$\xi_L=\xi / a=1/m_L$. Consider a simulation on a linear lattice of $N_1$ 
lattice point(in figure \ref{fig:scaling}, $N_1=8$). Suppose the simulation
with initial parameters \{$P_1$\} gives output correlation length in lattice
units $\xi_{L_1}$ (in figure \ref{fig:scaling}, $\xi_{L_1}=4$). In the
computer simulation all quantities are in lattice units; there is no physical
lattice spacing. We are free to introduce a lattice constant $a_1$ in some 
arbitrary physical units.  Then the correlation length in physical units is 
$\xi_1=\xi_{L_1}\times a_1$($=4a_1$). We now want to proceed toward the 
continuum limit $a_{\infty}\rightarrow 0$. We do this by a sequence of halving
the {\em physical} lattice spaceing, as follows:
Go to lattice of $N_2=2N_1$($=16$ in figure \ref{fig:scaling}) lattice
points. Adjust the input parameters \{$P_2$\} so that the output correlation
length in lattice units in $\xi_{L_2}=2\xi_{L_1}$($=8$). As the figure
shows, this may be interpreted a $a_2=a_1/2$ and 
$\xi_2=\xi_{L_2}\times a_2=\xi_1$, {\em i.e.} we have halved the {\em physical}
lattice spacing while keeping the physical volume and the correlation length 
in physical units fixed.

When $a$ goes to zero, we have
\begin{eqnarray}
a & \longrightarrow & 0 \\
\xi_L &\longrightarrow & \infty,\\
m_L & \longrightarrow & 0
\end{eqnarray}

 For the $\lambda$, we have two different possibilities\cite{FF:rwqft}:
\begin{description}
\item[(1)]  $\lambda~~fixed$, so that when $a\rightarrow 0$, 
$\lambda_L \rightarrow 0$. This is the standard procedure in 
superrenormalizable field theory($d < 4$). From the point of view of 
field theory, no nontrivial coupling-constant renormalization is being
performed; and none need be performed, since there is no ultraviolet 
divergences for the coupling constant renormalization in $d < 4$. 
From the point of view of statistical mechanics,
the theory is becoming extremely {\em weakly} coupled(
$\lambda_L \sim a^{4-d}$); but the effects of this coupling are amplified
by the {\em infrared} divergences of the $\phi^4$ lattice theory near the
critical point in $d < 4$, leading to a non-Gaussian continuum limit.

\item[(2)] $\lambda_L~~fixed$, so that when $a\rightarrow 0$,
$\lambda \rightarrow \infty$. This is the standard procedure in
the statistical mechanical theory of critical phenomena. In this case it
is for $d > 4$ that matters are simple: correlation function are given by
a perturbation expansion that is free of infrared divergences. From the
point of view of field theory this might seem surprising, since the 
$\phi^4_d$ field theory for $d > 4$ is perturbatively nonrenormalizable,
with horrendous ultraviolet divergence. However, it is mitigate by the fact
that the theory is becoming extremely {\em weakly} coupled(
$\lambda \sim a^{d-4}$). On the other hand, for $d < 4$ the theory of
critical phenomena is very complicated, by virtue of the infrared 
divergences. This is reflected in field theory in the fact that the theory
is becoming extremely {\em strongly} coupled($\lambda \sim a^{d-4}$), 
causing perturbation theory to break down even though the internal
momentum integrations are ultraviolet convergent.

\end{description}

\begin{figure}[tbp]
\begin{center}
\leavevmode
\epsfxsize 5in
\epsfbox{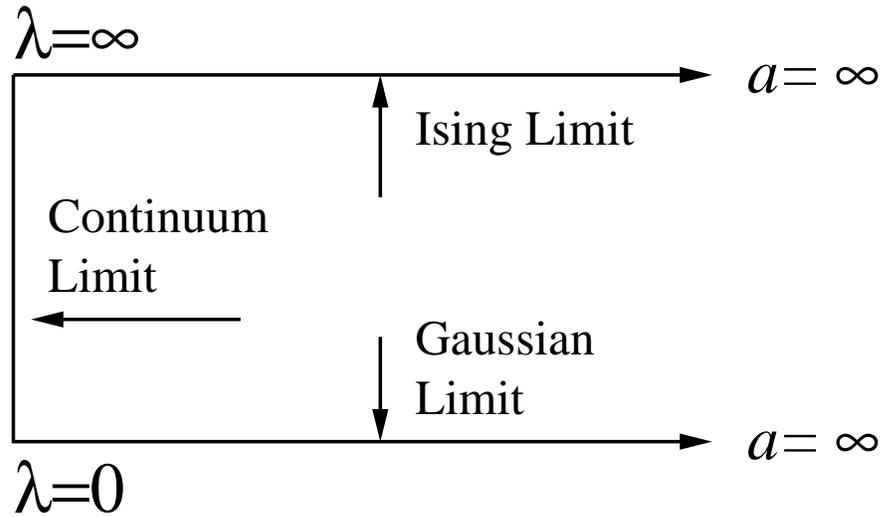}
\caption{The two parameters ($\lambda,a$) phase diagram for different limits}
\label{fig:conlim}
\end{center}
\end{figure}

\begin{figure}[p]
\begin{center}
\leavevmode
\epsfxsize 5in
\epsfbox{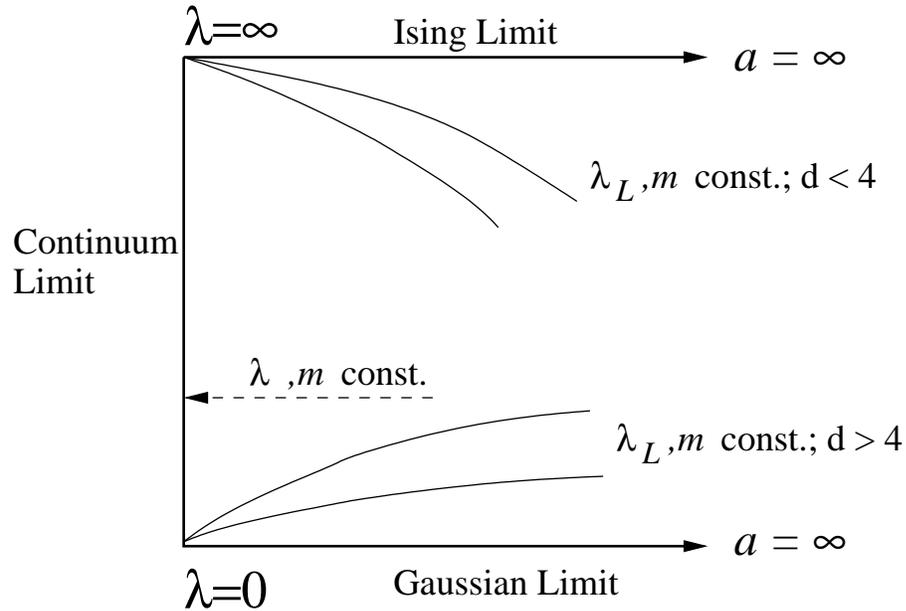}
\caption{The two parameters ($\lambda,a$) phase diagram. The diagram shows the
statistical mechanical approach for both $d<4$ and $d>4$. The dashed arrow 
shows the field theory approach.}
\label{fig:conlim2}
\end{center}
\end{figure}

The Figure \ref{fig:conlim} shows the different limits in the two parameter 
($\lambda,a$) plane. When $\lambda \rightarrow \infty$, that corresponds to the
Ising limit. When $\lambda \rightarrow 0$, that corresponds to the Gaussian 
limit.
When $a \rightarrow 0$, that corresponds to the continuum limit. The Figure
\ref{fig:conlim2} shows the statistical mechanical approaches for $d < 4$ and
$d>4$(the solid lines), also for the field theory approach(the dashed arrow).

Since we want to consider a superrenormalizable theory with fixed weak coupling
constant, we use the first alternative. Hence when $a$ goes to zero,
the $\lambda_L$ should also goes to zero.
\begin{equation}
a\longrightarrow 0,~~~~~\lambda_L \longrightarrow 0
\end{equation}
In the simulation, when we halve the lattice constant $a\rightarrow a/2$, 
we should have $\lambda_L \rightarrow \lambda_L /2$ and 
$\xi_L \rightarrow 2\xi_L$. 

\begin{figure}[p]
\begin{center}
\leavevmode
\epsfxsize 6in
\epsfbox{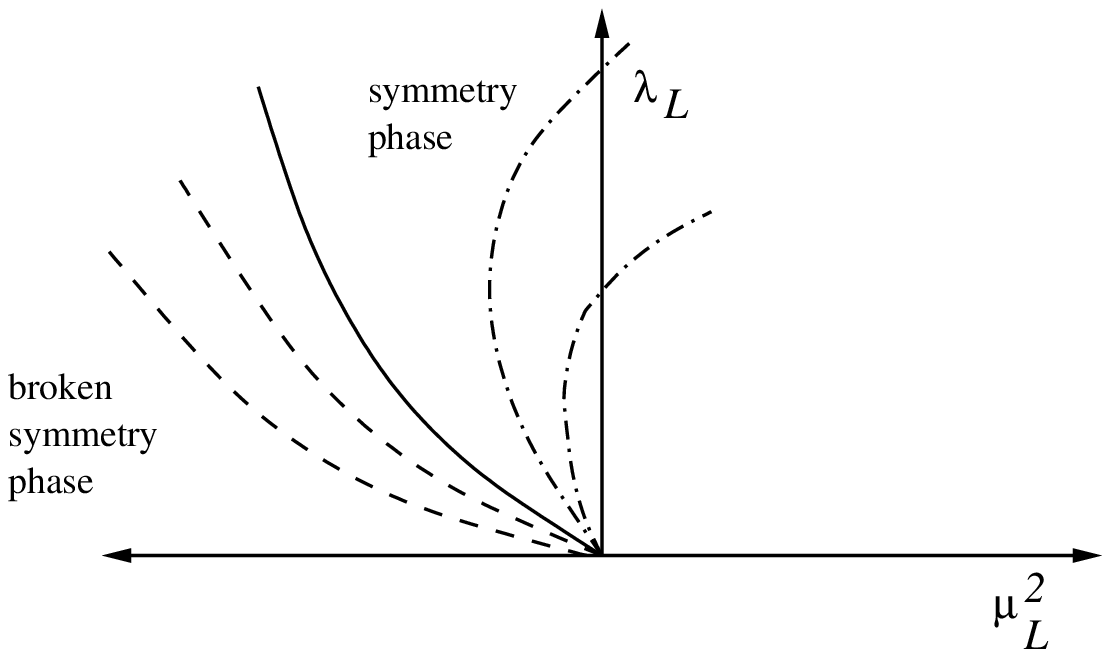}
\caption{The phase diagram for the three-dimensional, single component scalar 
$\phi^4$ theory. The solid line separates the symmetric phase from the broken 
symmetry phase. The dash-dot lines represent lines of constant physics in the
symmetry phase, and the dashed lines are lines of constant physics in the 
broken phase. The origin is the Gaussian limit and does not imply a trivial 
field theory for $d<4$.}
\label{figure:phasediag}
\end{center}
\end{figure}

The phase diagram of the theory in terms of the parameters $\mu^2_L$ and 
$\lambda_L$ is sketched in Figure \ref{figure:phasediag}. 
The Gaussian fixed point at 
$(\mu^2_L,\lambda_L)=(0,0)$ is not trivial in less than four dimensions,
as has been demonstrated by constructive field theory 
techniques\cite{GJ:qphy,FF:rwqft}.
The physical dimensionful coupling constant remains finite as this
Gaussian fixed point is approached along the dashed or dotted lines sketched
in Figure \ref{figure:phasediag}. The solid line extending outward from the 
origin is the line of 
the second-order phase transition separating the broken symmetry phase
from the symmetric phase. In this space of the lattice coupling constant
and lattice bare mass squared, continuum physics is recovered by moving
toward the origin along line determined by the renormalization group flow. 
Along these lines the ratio of the renormalized coupling and mass are
constant. We want to emphasize that we are not working near the Ising
fixed point, which in this parameterization is the point on the solid line
with $\lambda_L \rightarrow \infty$.

Other dimensionless properties \{$P$\} of the system should also remain 
fixed when we approach to the continuum limit. One expects to approach the 
continuum limit
for $\xi_L \gg 1$. If one starts far from continuum limit($\xi_L$ not $\gg 1$),
the \{$P$\} will not stay fixed(within statistical errors) until after a number
of steps have got to large enough $\xi_L$(see figure \ref{fig:scaling}). 
When this has been achieved,
one claims that one is so close to the continuum limit values of the
\{$P$\} that error due to lattice contamination is less than the statistical
error of the simulation.

\section{The Simulation Methods}

Here we describe step by step how we carry out the simulation and analysis
the data. 
Because we know the continuum limit of $1/Z'$ and $m'$ exists
\cite{GJ:qphy,FF:rwqft}, we will try
to halve the lattice constant $a$ until those values settle to their
continuum limits(invariant under $a \rightarrow a/2$). The method is:
\begin{description}
\item[(1).] Choose the proper values for $\lambda_L$ and $\mu^2_{0_L}$ as 
the input parameters and calculate the $1/Z'$ and $m'^2$ by measuring the 
slope and intercept of the $G^{-1}$ plot in section 5.1. 
\item[(2).] Then we check whether the correlation length
$\xi_L$ satisfying the conditions in equation \ref{xi:cond} or not. 
\item[(3).] If it satisfies the condition, then the value of $1/Z'$ will
gives us the upper bound of inclusive cross section.
\item[(4).] If not, we halve our lattice space(go to higher energy but fix 
the ratio of $E/n$) and, according to the continuum 
limit, move to the the next point on the phase diagram along the dash line
in Figure \ref{figure:phasediag}.
In our simulation, halving the lattice constant and keep the volume fixed
is the same as keeping the lattice constant fixed and double the lattice
size. So we have $N \rightarrow 2N$ and set 
$\lambda_L \rightarrow \lambda_L /2$ then find the proper value of 
$\mu^2_{0_L}$ (actually, $\mu^2_{0_L} \rightarrow \mu^2_{0_L} /2$ 
approximately)\footnote{This is because in 3-D $\phi^4$ theory, the bare mass
square, $\mu^2_{0_L}$, is linearly UV divergent and propotional to $m_L$ 
and when $a \rightarrow a/2$, $m_L \rightarrow m_L /2$.} 
to makes  $\xi_L \rightarrow 2\xi_L$. Under this condition, we keep the 
dimensionless
quantity $\lambda / m'$ the same. This step will guarantee we
stay on the dash line in Figure \ref{figure:phasediag} but moves close to the 
origin. 
\item[(5).] At this new scale, measure the $1/Z'$ and $m'$ and check
the condition \ref{xi:cond} again. If the correlation length still can not
satisfy the condition or it satisfies but the values $1/Z'$ and $m'$ do not
remain the same, we then repeat step (4) until the 
correlation length can satisfy the equation \ref{xi:cond} and both 
$1/Z'$ and $m'$ do not change anymore.(stay within estimated errors bounds)

\end{description}

The last step checks the continuum limit of the simulation. 
If $1/Z'$ and $m'$ are
insensitive to the lattice space, the value will be very close to the 
continuum limit(with some statistical errors).
In the real life, we can not keep enlarging the lattice size because of the 
limit of our computing power. We can only choose the starting point by 
try and error at small lattice size $N=32$, then move up to the maximum 
lattice size $N=256$.

\section{Verify the Ising limit of the Simulation}

From Figure \ref{fig:conlim}, one can see that if the $\lambda$ in the
$\phi^4$ theory goes to infinity, it becomes the Ising theory. The Ising
model Monte Carlo simulation have been well known in statistical mechanics
\cite{tsypin,Jasnow}.  
To verify our Monte Carlo simulation in three dimension $\phi^4$ theory, 
one can compared the results of three dimensional $\phi^4$ at Ising limit
($\lambda \rightarrow \infty$) with the results from Ising model simulation.
It is well known that the critical temperature $\beta_c$  for the three 
dimensional Ising model is about $\beta_c = 0.2217$. The $\phi^4$ theory
should be able to reproduce this result by letting the 
$\lambda \rightarrow \infty$. We can rewrite the field theory Lagrangian 
\ref{disLarg} as
\begin{equation}
{\cal L} = -\sum_{\vec{i}\neq \vec{j}}\phi_{\vec{i}} \phi_{\vec{j}} + 
\sum_{\vec{i}}\frac{\lambda}4 \left( \phi^2_{\vec{i}} - 
\frac{-\mu^2_L+2d}{\lambda} \right)^2  \label{eq:ftising}
\end{equation}
where the $d$ is the dimension. When $\lambda \rightarrow \infty$, it should 
goes back to the Ising model. Compare with the Hamiltonian of Ising model
\begin{equation}
{\cal H} = -\beta \sum_i \sigma_i\sigma_j  \label{IsingH},
\end{equation}
we see that the $\lambda \rightarrow \infty$ will force the second term in 
equation \ref{eq:ftising} to zero. 
So, we can have
\begin{equation}
\lim_{\lambda\rightarrow \infty} \phi^2_{\vec{i}} - \frac{-\mu^2_L+2d}{\lambda} =0
\end{equation}
This suggest that 
\begin{equation}
\lim_{\lambda \rightarrow \infty} \phi_{\vec{i}} = \pm 
\sqrt{\frac{-\mu^2_L+2d}{\lambda}} \equiv \sqrt{\beta}\sigma_{\vec{i}}
\end{equation}
where the $\sigma_{\vec{i}}= \pm 1$ and 
\begin{equation}
\beta = \frac{-\mu^2_L+2d}{\lambda}
\end{equation}
Under this limit, the Hamiltonian of $\phi^4$ theory becomes
\begin{eqnarray}
{\cal H} & = & \lim_{\lambda \rightarrow \infty} 
\sum_{\vec{i}\neq \vec{j}}\phi_{\vec{i}} \phi_{\vec{j}} -
\sum_{\vec{i}}\frac{\lambda}4 \left( \phi^2_{\vec{i}} - 
\frac{-\mu^2_L+2d}{\lambda} \right)^2  \nonumber \\
& = & -\beta\sum_{\vec{i}\neq \vec{j}} \sigma_{\vec{i}} \sigma_{\vec{j}} 
\end{eqnarray}
which is the Ising model with the external field $J=0$.

To verify the Ising limit, 
we do the simulation with the $\lambda \gg 1$  and
calculate the VEV $\langle \phi \rangle$ for different $\mu^2_0$. We start
in the symmetric phase with one $\lambda$ value and change the value $\mu^2_0$ 
until the VEV is no longer zero.   In our simulation, when $\lambda \gg 1$, we
get a very sharp transition from the symmetric phase to broken symmetry phase.
So, it is very easy to identify the critical $\mu^2_{0_c}$.
Table \ref{isinglim} shows the $\beta_c$ from the simulation at $32^3$ 
with different
$\lambda$. The third row shows that when $\lambda=1000$, the $\beta_c$ is 
very close to the value 0.2217. This give us confidence in our 
simulation.

\begin{table}[tbp]
\caption{The Monte Carlo simulation of $32^3$ $\phi^4_3$ theory at Ising 
limit($\lambda \rightarrow \infty$). The table shows the critical temperature
$\beta_c$ at different $\lambda$ values.}
\label{isinglim}
\begin{center}
\begin{tabular}{|c|c|c|c|}
\hline
$\lambda$ & $\mu^2_{0_c}$ & $-\frac{\mu^2_{0_c}}{\lambda}$ & $\beta_c$ \\
\hline
100 & -31 & .31 & .25 \\
300 & -75 & .25 & .23 \\
1000 & -228 & .228 & .222 \\
\hline
\end{tabular}
\end{center}
\end{table}

\chapter{The Perturbation Calculation}

\section{$\phi^4_3$ model and renormalization}

\begin{figure}[tbp]
\begin{center}
\leavevmode
\epsfxsize 4in
\epsfbox{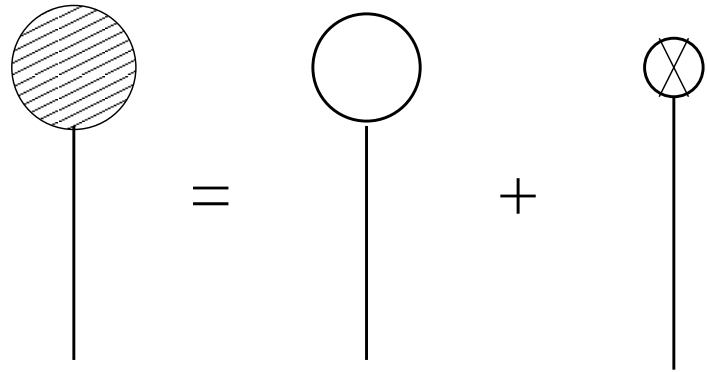}
\caption{The one-loop tadpole diagram and the counter term.}
\label{fig:tadpole}
\end{center}
\end{figure}

To compare with the simulation results, we have done the perturbation 
calculation for the VEV, $\langle \phi\rangle / \sqrt{m'}$, and $1/Z'$ 
in three dimensional $\phi^4$ theory. We use the standard Lagrangian
\begin{equation}
{\cal L} = \frac12 \partial\phi_0(x) \partial\phi_0(x) - 
\frac{\mu^2_0}{2} \phi^2_0(x) - \frac{\lambda_0}{4}\phi^4_0(x)
\end{equation}
The field are unrenormalized canonical field, and $\mu^2_0$, $\lambda_0$ are
bare mass and coupling constants. To implement perturbation theory in the
broken symmetry phase($\mu^2_0 < 0, \langle \phi \rangle \neq 0$), the
field is shifted
\begin{equation}
\langle \phi_0 \rangle = {\cal V}_0,~~~~~~\phi_0 = {\cal V}_0 + {\hat \phi}_0
\end{equation}
The condition that one is perturbing about the correct vacuum is
\begin{equation}
\langle {\hat \phi}_0 \rangle = 0  \label{vev:0}
\end{equation}
This condition determines ${\cal V}$ as a function of the parameters of the
theory. The bare field and parameters are renormalized  as
\begin{equation}
\phi_0 = \sqrt{Z}\phi,~~~~~~\mu^2_0=Z_{m}\mu^2, ~~~~~~{\cal V}=\sqrt{Z}V
\end{equation}
As we mentioned before, in three dimension super renormalizable theory, 
there is no need to renormalize the coupling constant $\lambda$. Hence we can 
set
\begin{equation}
\lambda_0 = \lambda,~~~~~Z_{\lambda}=1
\end{equation}
When we substitute the renormalized field in to the bare Lagrangian, and write 
all $Z$'s as $Z=1+\delta Z$, one can get a renormalized Lagrangian plus
terms proportional to $V$(and containing no $\delta Z$), plus counter terms
which are proportional to one or more $\delta Z$'s. 
To implement the stability condition
\ref{vev:0}, compute the tadpole graphs, including counter terms, to a given
order(number of loops) in perturbation theory, and expand the renormalized V
\begin{eqnarray}
V & \equiv & \zeta_v v, \\
\zeta_v & = & 1 + \zeta^{(1)}_v + \zeta^{(2)}_v + \cdots
\end{eqnarray}
and adjust $\zeta_v$(equivalently $V$) to satisfy \ref{vev:0} order by order. 
In the broken symmetry phase, we can expand the linear $\hat\phi$ term in the 
Lagrangian(up to one loop order)
\begin{eqnarray}
{\cal L} & = & -(\mu^2_0 + \lambda V^2)V\hat{\phi}_0 \\ 
& = & -(Z_m\mu^2 + Z\zeta^2_v \lambda v^2)Z\zeta_v v \hat{\phi}  \\
&=&-(\mu^2 + \lambda v^2)v\hat\phi 
-(\mu^2 + \lambda v^2)(\zeta_v -1)v\hat{\phi} \nonumber \\
&& -[(Z_mZ-1)\mu^2 + (Z^2\zeta^2_v-1)\lambda v^2]v\hat{\phi}/ \\
&+& \mbox{ two-loop counter terms}
\end{eqnarray}
The zeroth order fixes
\begin{equation}
v^2=-\frac{\mu^2}{\lambda},~~~~~~(\mu^2 = -\frac{m^2}{2} < 0)
\end{equation}
and the one loop counter term may be written as
\begin{eqnarray}
{\cal L}_1 & = & v\delta\mu^2 \hat{\phi}  \\
\delta\mu^2 & =& (Z_mZ-Z^2\zeta^2_v)\lambda v^2 + \mbox{two-loops term}
\end{eqnarray}
Computation of $\langle {\hat \phi} \rangle$ in one-loop order(Figure 
\ref{fig:tadpole}) give
\begin{equation}
\langle {\hat \phi} \rangle = -i3\lambda v A_{m} + i v\delta\mu^2 = 0
\end{equation}
where the regularized Feynman integral is
\begin{equation}
A_m = i \int \frac{d^3p}{(2\pi)^3} \frac1{p^2-m^2+i\epsilon}
\end{equation}

The $\hat{\phi}^2$ terms in the Lagrangian gives
\begin{equation}
{\cal L}_2 = \partial \phi\partial \phi - \frac{m^2}{2}\hat{\phi}^2 + 
\frac12 \delta\mu^2\hat{\phi}^2 - \frac{m^2}2(Z^2\zeta^2_v-1)\hat{\phi}^2  
\label{ct:phi}
\end{equation}
the one-loop counter terms(for $ -i\Sigma$) generated by equation \ref{ct:phi}
are
\begin{equation} 
CT = i(Z-1)q^2 + i\delta\mu^2 - i(Z^2\zeta^2_v-1)m^2
\end{equation}
The one-loop self-energy diagrams are shown in Fig. 11. The tadpole diagram 
plus the tadpole counter term add to zero by our previous choice of 
$\delta\mu^2$. Hence, the renormalized self-energy function is
\begin{equation}
-i\Sigma(q^2)=-i18\lambda^2 v^2 I_{mm}(q^2)+ i\delta Zq^2-i(Z^2\zeta^2_v-1)m^2
\label{sig:q}
\end{equation}

\begin{figure}[tbp]
\begin{center}
\leavevmode
\epsfxsize 6in
\epsfbox{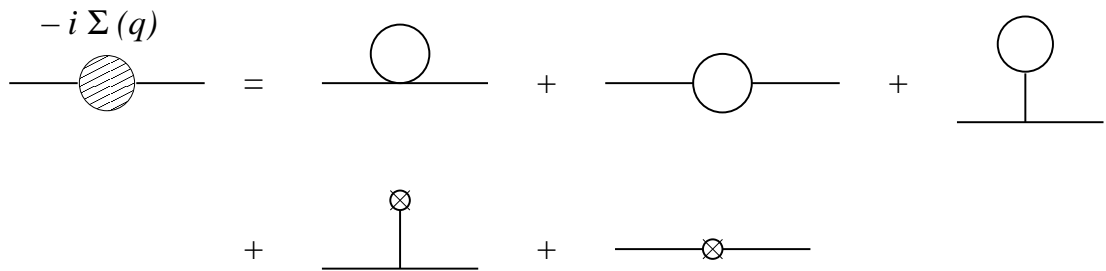}
\caption{The one-loop self energy diagrams.}
\end{center}
\end{figure}

To determine the renormalization constants, we can impose the on-shell 
renormalization conditions
\begin{equation}
\Sigma(q^2=m^2) = 0, \left. \frac{\partial}{\partial q^2}\Sigma(q^2) 
\right|_{q^2=m^2} = 0
\end{equation}
This give us the on-shell renormalization constants and $\delta\zeta_v$
\begin{eqnarray}
\delta Z &=& \frac{9\lambda}{4\pi m} \left( -\frac13 + \frac12 \ln\sqrt{3} 
\right)  \label{dZ}\\
\delta\zeta_v &=& \frac{9\lambda}{4\pi m}\left( \frac16 + \frac14 \ln\sqrt{3}
\right) \\
\delta Z_m &=& \frac{9\lambda}{4\pi m}\left( \frac2{3m^2} + \ln\sqrt{3} 
\right)
\end{eqnarray}
Alternatively, the $Z'$ and $m'$ are defined by the zero momentum 
renormalization conditions
\begin{equation}
\Sigma(q^2=0) = 0,~~~~~~ \left. \frac{\partial}{\partial q^2}\Sigma(q^2)
\right|_{q^2=0} = 0 \label{zeroprc}
\end{equation}
which give
\begin{eqnarray}
\delta Z' &=& -\frac{3\lambda}{32\pi m}  \\
\delta\zeta_v ' &=& \frac{21\lambda}{32\pi m} \\
\delta Z'_m &=& \frac{39\lambda}{32\pi m} + \frac{3\lambda}{2\pi m^3}
\end{eqnarray}

\section{One-Loop calculation for $\frac{\langle \phi_0 \rangle}{\sqrt{m'}}$ }

We calculate the vacuum expectation value(VEV) to one-loop order and compare
to the simulation results.
In the simulation, we measure the bare VEV and $1/Z'$ and $m'$. 
In three dimensions, $\langle \phi_0 \rangle$ has dimension 
$\sqrt{\mbox{mass}}$.
Hence, the quantity $\langle \phi_0 \rangle /\sqrt{m'}$ is dimensionless and
we can look for its continuum limit.
In the broken symmetry phase, the VEV will is
\begin{equation}
\langle \phi_0 \rangle = V'_0 = \sqrt{Z'}\zeta'_v v',~~~~~~~v'=\frac{m'}{\sqrt{2\lambda}}
\end{equation}
So we have
\begin{eqnarray}
\frac{\langle \phi_0 \rangle}{\sqrt{m'}} & = &\sqrt{\frac{m'}{2\lambda}}\left(
1+\frac12\delta Z' + \delta\zeta'_v  \right) \\ 
& =&  \sqrt{\frac{m'}{2\lambda}} \left(
1 + \frac{9\lambda}{4\pi m}\frac{13}{48} \right) \label{vev}
\end{eqnarray}

\section{The calculation for $\frac1{Z'}$ to $O(\lambda^2$)}

In chapter 3(and Appendix A), we have derived a bound on the inclusive cross 
section. 
\begin{equation}
\int ds~\sigma(s) \le \frac1{Z'} -1 \label{eq:cros}
\end{equation}
The lattice Monte Carlo simulation p[rovides a nonperturbative result for the 
right hand side of equation \ref{eq:cros}. To see if this implies any large 
nonperturbativecontribution, we have to compare it to a perturbative case of 
the equality.(The ${\cal C}$ introduced in appendix A is zero in the weak 
coupling limit.)

\begin{equation}
Z\left( \frac1{Z'} - 1  \right) = \int ds ~\sigma(s) \equiv  {\cal X}
\equiv \sum_{N} {\cal X}^{(N)}
\end{equation}
The index $(N)$ indicate the order of $\lambda^N$. So that
\begin{equation}
\frac1{Z'} -1 = \frac{\cal X}{Z} = {\cal X}^{(1)} - \delta Z {\cal X}^{(1)} +
{\cal X}^{(2)} + \cdots \label{XZ}
\end{equation}
where 
\begin{equation}
{\cal X}^{(N)} = \sum_n \int ds ~\sigma^{(N)}_n
\end{equation}
Here we have to sum over all final states with $n$ outgoing particles. 
\begin{equation}
\sigma_n(s) = \frac{1}{2\pi s^2}\frac1{n!} \int d\Phi_n(s) 
|\underline{G}\hspace{-3mm}/~^c_{n+1}(p_1,\dots,p_n,P)|^2
\end{equation}
where the phase space integral is
\begin{equation}
\int d\Phi_n = \frac1{n!}\int \frac{d^3p_1}{(2\pi)^3}\delta(p^2_1-m^2)\cdots
\int \frac{d^3p_n}{(2\pi)^3}\delta(p^2_n-m^2)
(2\pi)^3\delta^3(\vec{P}-\sum p_i)
\end{equation}
The lowest order, $O(\lambda)$, contribution has only one diagram. Figure 
\ref{fig:tree}
shows the lowest order tree diagrams. 
So that
\begin{eqnarray}
{\cal X}^{(1)} & = & \int ds \sigma^{(1)}_{2} \nonumber \\
& = & \int~ds \frac1{2\pi s^2} \frac1{2!} \int d\Phi_2 
|-i3\sqrt{2}m\sqrt{\lambda}|^2 \nonumber \\
& = & \frac{9 m^2 \lambda}{2\pi s^2} \int d\Phi_2
\end{eqnarray}
If we choose the CM frame, the two particle phase space integral will be
\begin{equation}
\int d\Phi_2 = \frac1{4\sqrt{s}} \label{phI:2}
\end{equation}
Hence the contribution of order $O(\lambda)$ is
\begin{equation}
{\cal X}^{(1)} = \frac{3\lambda}{32\pi m} \label{X(1)}
\end{equation}
By using equation \ref{dZ}, the second term in equation \ref{XZ} will be(in
order of $O(\lambda^2)$)
\begin{equation}
-\delta Z^{(1)}{\cal X}^{(1)} = -\left( \frac{3\lambda}{32\pi m} \right)^2 
\left( -8 + 6\ln 3 \right)  \label{dzX(1)}
\end{equation}

\begin{figure}[tbp]
\begin{center}
\leavevmode
\epsfxsize 6in
\epsfbox{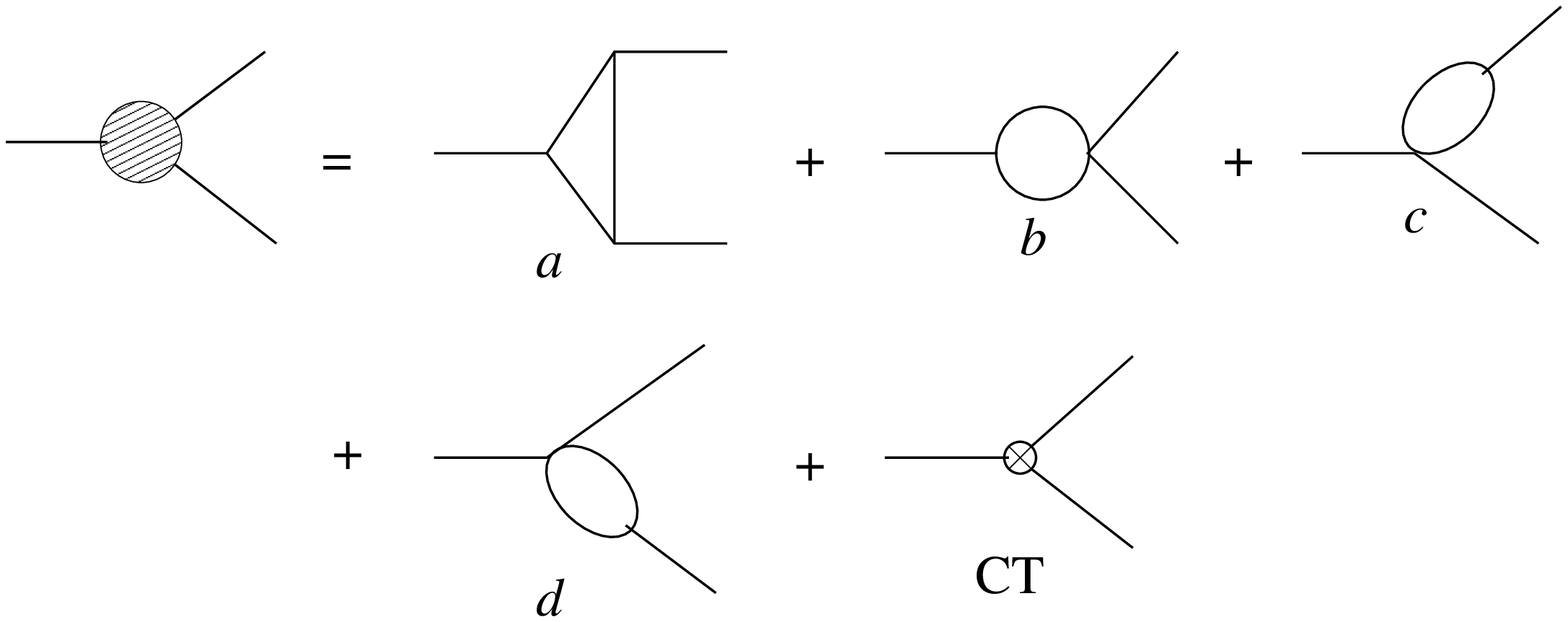}
\caption{The one-loop Feynman diagrams for $n=2$($G\hspace{-3mm}/~^{\frac32}_3$).}
\label{fig:oneloop}
\end{center}
\end{figure}

\begin{figure}[tbp]
\begin{center}
\leavevmode
\epsfxsize 6in
\epsfbox{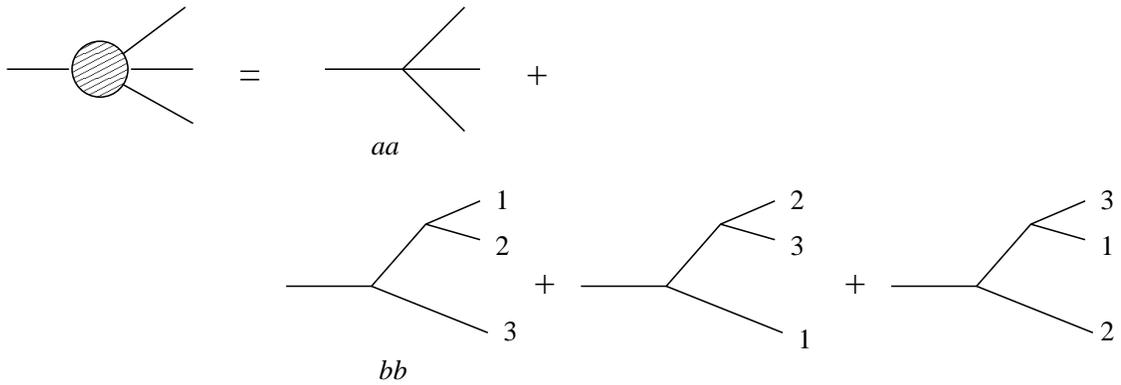}
\caption{The tree diagrams for $n=3$($G\hspace{-3mm}/~^2_2$).}
\label{fig:tree}
\end{center}
\end{figure}

The Feynman diagrams and the counter terms for the third term in 
equation \ref{XZ} are shown in Figure \ref{fig:oneloop} and Figure 
\ref{fig:tree}. Figure \ref{fig:oneloop} shows the
one-loop Feynman diagram with $n=2$($G\space{-3mm}/^{\frac32}_3$) and Figure 
\ref{fig:tree} shows 
the tree level Feynman diagram for $n=3$($G\hspace{-3mm}/~^{\frac32}_4$). 
The cross section
of order $\lambda^2$ is
\begin{eqnarray}
\sigma^{(2)}_2(s) & = & \frac1{2\pi s^2} \frac12 \int \Phi_2 
2\Re [G\hspace{-3mm}/~^{\frac12}_3 G\hspace{-3mm}/~^{\frac32}_3 ]  \\
\sigma^{(2)}_3(s) & = & \frac1{2\pi s^2} \frac1{3!} \int \Phi_3 
2\Re [G\hspace{-3mm}/~^{\frac12}_4 G\hspace{-3mm}/~^{\frac32}_4 ] 
\end{eqnarray}
We can write
\begin{equation}
{\cal X}^{(2)} = \int ds~(\sigma^{(2)}_2 + \sigma^{(2)}_3) = {\cal X}^{(2)}_2 +
{\cal X}^{(2)}_3
\end{equation}
The one-loop Feynman integrals(with $r=\frac{s}{m^2}$) in Figure
\ref{fig:oneloop} are 
\begin{eqnarray}
I_{a}(s) & = & i\frac{9\sqrt{2}\lambda^{\frac32}}{8\pi} 
\left( -\frac6{2\sqrt{r-3}} \right)  \nonumber \\
&\times& \left\{ 
\frac{x_2}{y_2}\left[ \ln\left(\frac{-1-x_2-2y_2}{(-1+x_2)x_2y_2} \right) -
\ln\left(\frac{-2+x_2 -2y_2}{x^2_2y_2}\right) \right]  \right.  \nonumber \\
&& -\left. \frac{x_1}{y_1}                
\left[ \ln\left(\frac{-1-x_1-2y_1}{(-1+x_1)x_1y_1} \right) -
\ln\left(\frac{-2+x_1 -2y_1}{x^2_1y_1}\right) \right] \right\} \\
I_{b}(s) & = & i\frac{9\sqrt{2}\lambda^{\frac32}}{8\pi} \times
\frac{2m}{\sqrt{s}}\left[ \ln\left(\frac{\sqrt{s} + 2m}{\sqrt{s}-2m}\right) 
\right] \\
I_{c}(s) &=& i\frac{9\sqrt{2}\lambda^{\frac32}}{8\pi} \times 2\ln3 \\
I_{d}(s) &=& i\frac{9\sqrt{2}\lambda^{\frac32}}{8\pi} \times 2\ln3 = I_{c} \\
\mbox{C.T.} &=& i\frac{9\sqrt{2}\lambda^{\frac32}}{8\pi} \left[ 3-\frac{15}{4} 
\ln 3 \right]
\end{eqnarray}
where $x_1$, $x_2$, and $y_i$ are
\begin{eqnarray}
x_1 &=& \frac{2+2\sqrt{r-3}}{4-r} \\
x_2 &=& \frac{2-2\sqrt{r-3}}{4-r} \\
y_i &=& \sqrt{x^2_i - x_i +1}
\end{eqnarray}
The two particle phase space integral is the same as before(equation 
\ref{phI:2}). Finally, after calculating the $s$ integral, one gets
\begin{eqnarray}
{\cal X}^{(2)}_2 &=& \int^{\infty}_{4m^2}ds~\sigma^{(2)}_2 \\
&=&  -\left( \frac{3\lambda}{8\pi m} \right)^2 \times 0.496728 \label{X(2)2}
\end{eqnarray}

The Figure \ref{fig:tree} shows the $n=3$ case of $O(\lambda^2)$. 
\begin{equation}
{\cal X}^{(2)}_3 = \int ds~\sigma^{(2)}_3
\end{equation}
and
\begin{equation}
\sigma^{(2)}_3 = \frac1{2\pi s^2}\frac1{3!} \int d\Phi_3~|G\hspace{-3mm}/~_4|^2
\end{equation}
From Figure \ref{fig:tree}, the four point function is
\begin{equation}
G\hspace{-3mm}/~_4 = T_{aa} + T_{bb}
\end{equation}
the $T_{aa}$ is just the tree diagram in Figure \ref{fig:tree}$aa$.
\begin{equation}
T_{aa} = -i6\lambda
\end{equation}
The tree level diagram in Figure \ref{fig:tree}$bb$ has six permutations. 
\begin{equation}
T_{bb} = -i 18m^2 \lambda \left[\frac1{s_{12} - m^2} +\frac1{s_{23} - m^2} +
\frac1{s_{31} - m^2} \right]
\end{equation}
where 
\begin{equation}
s_{ij} = (k_i+k_j)^2
\end{equation}
is the sum of the momentum of two of the final particles. 
We make the same simplification that was made on the calculations of the 
multipicle phase space {\it ie.} we replace each $k_i\cdot k_j$ by the average 
$\langle k_i\cdot k_j \rangle$. Then
\begin{equation}
\langle s_{12} - m^2 \rangle = \langle s_{23} - m^2 \rangle = \langle s_{31} - m^2 \rangle = \frac{s}3
\end{equation}
Hence, the four point function square can be simplified to 
\begin{equation}
|G\hspace{-3mm}/~^2_4(s)|^2 = 36\lambda^2 \left( 1+ 54\frac{m^2}{s} + 729\frac{m^4}{s}\right)
\end{equation}
Since the four point function does no longer depend on the individual momentum
of each final particle, the phase space integral becomes easy.
We can use the recursion relation \ref{PS:Rn} in chapter four to calculate the
space space integral. We find
\begin{equation}
\int d\Phi_3 = \frac1{16\pi} \left( 1-\frac{3m}{\sqrt{s}} \right)
\end{equation}
So the contribution from ${\cal X}^{(2)}_3$ becomes
\begin{eqnarray}
{\cal X}^{(2)}_3 &=& \int ds~\frac1{2\pi s^2}\frac16\int d\phi_3 |G\hspace{-3mm}/~^2_4(s)|^2\\
&=& \left( \frac{3\lambda}{8\pi m} \right)^2 \times 0.201276 \label{X(2)3}
\end{eqnarray}
Combining equation \ref{X(1)},\ref{dzX(1)},\ref{X(2)2}, and \ref{X(2)3}
gives
\begin{equation}
\frac1{Z'} = 1 + \frac{3\lambda}{8\pi m}\times \frac14 - 
\left( \frac{3\lambda}{8\pi m} \right)^2 \times 0.206944 \label{1Z'm}
\end{equation}

The mass $m$ in above equation is physical mass under the on-shell scheme. 
However, we only measure $m'$ from Monte Carlo simulation. In order to 
compare with the Monte Carlo results, one should use $m'$ in the perturbation
series. 
The relation between $m'$ and $m$ can be calculated from two point 
function.
\begin{equation}
G_0(p^2)=Z'G'(p^2)=ZG(p^2),
\end{equation}
and
\begin{equation}
G^{-1}_0(p^2=0)=\frac{m'^2}{Z'} = \frac1{Z}(m^2+\Sigma(p^2=0))
\end{equation}
So up to one-loop order, we have
\begin{eqnarray}
\frac{m'^2}{m^2} & = & \frac{Z'}{Z}(1+\frac1{m^2}\Sigma(p^2=0)) \\
& = & 1+\delta Z' - \delta Z + \frac1{m^2}\Sigma(0)  \\
& = & 1+\Delta  \label{mmratio}
\end{eqnarray}
From equation \ref{mmratio}, we have
\begin{eqnarray}
m &=& m' \left( 1 - \frac{\Delta}2 \right) \\
\Delta &=& \frac{9\lambda}{4\pi m} \left( -\frac{13}{24} + 
\frac12 \ln3 \right)
\end{eqnarray}
Plug into equation \ref{1Z'm}, we get 
\begin{equation}
\frac1{Z'} = 1 + \frac{3\lambda}{8\pi m'}\times \frac14 -
\left( \frac{3\lambda}{8\pi m'} \right)^2 \times 0.201214  \label{1Z'm'} 
\end{equation}

\chapter{Analysis and Results}

In this chapter, I will discuss the results from the Monte Carlo simulation
and compare the results with the perturbation results.

\section{The Results From the Simulation}

\begin{figure}[tbp]
\begin{center}
\vskip 1cm
\leavevmode
\epsfxsize 6in
\epsfbox{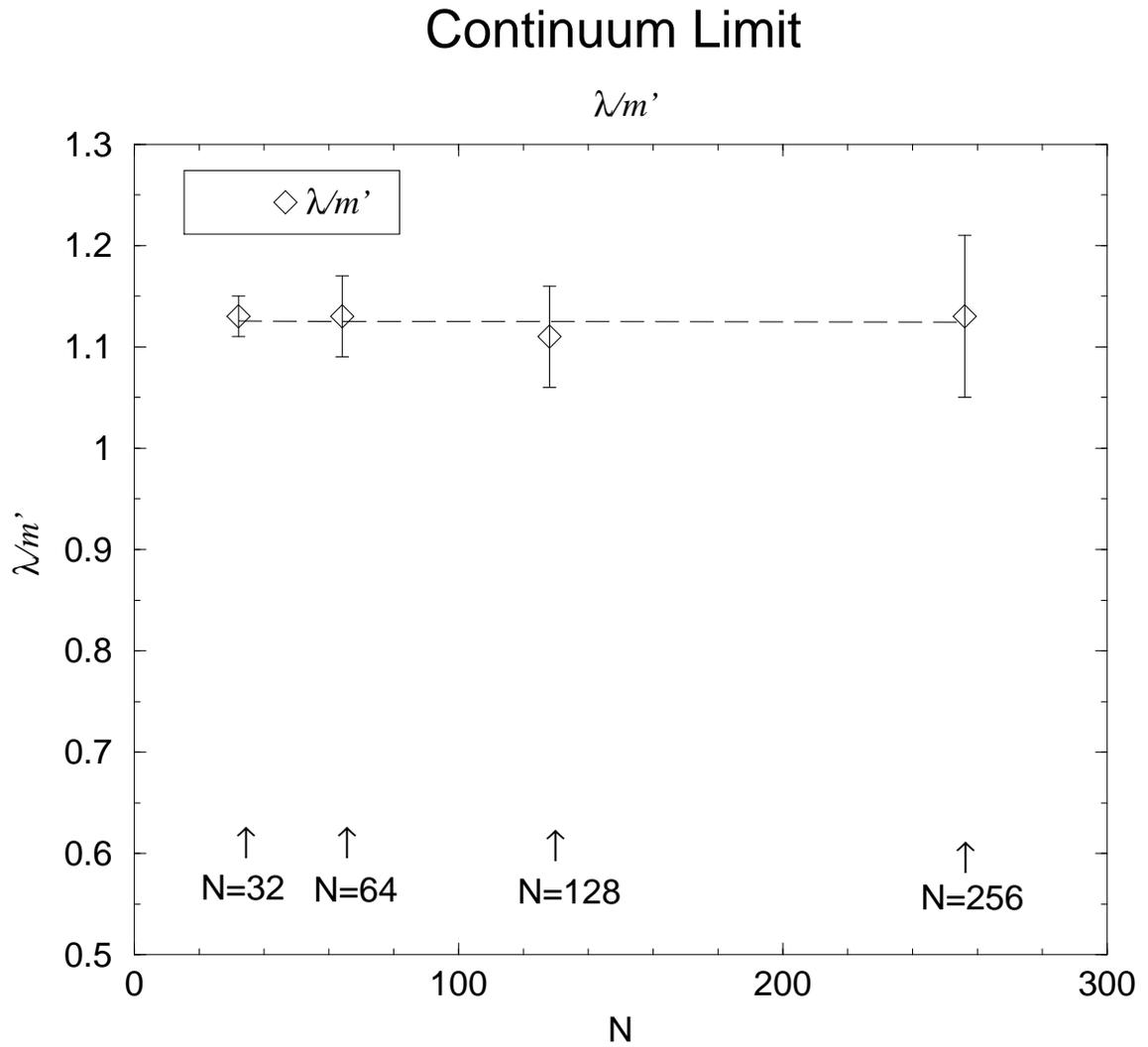}
\vskip 1cm
\caption{Continuum limit for the $\frac{\lambda}{m'}$. The
plot shows the insensitivity of lattice size for $\frac{\lambda}{m'}$.}
\label{fig:71}
\end{center}
\end{figure}

\begin{figure}[tbp]
\begin{center}
\vskip 1cm
\leavevmode
\epsfxsize 6in
\epsfbox{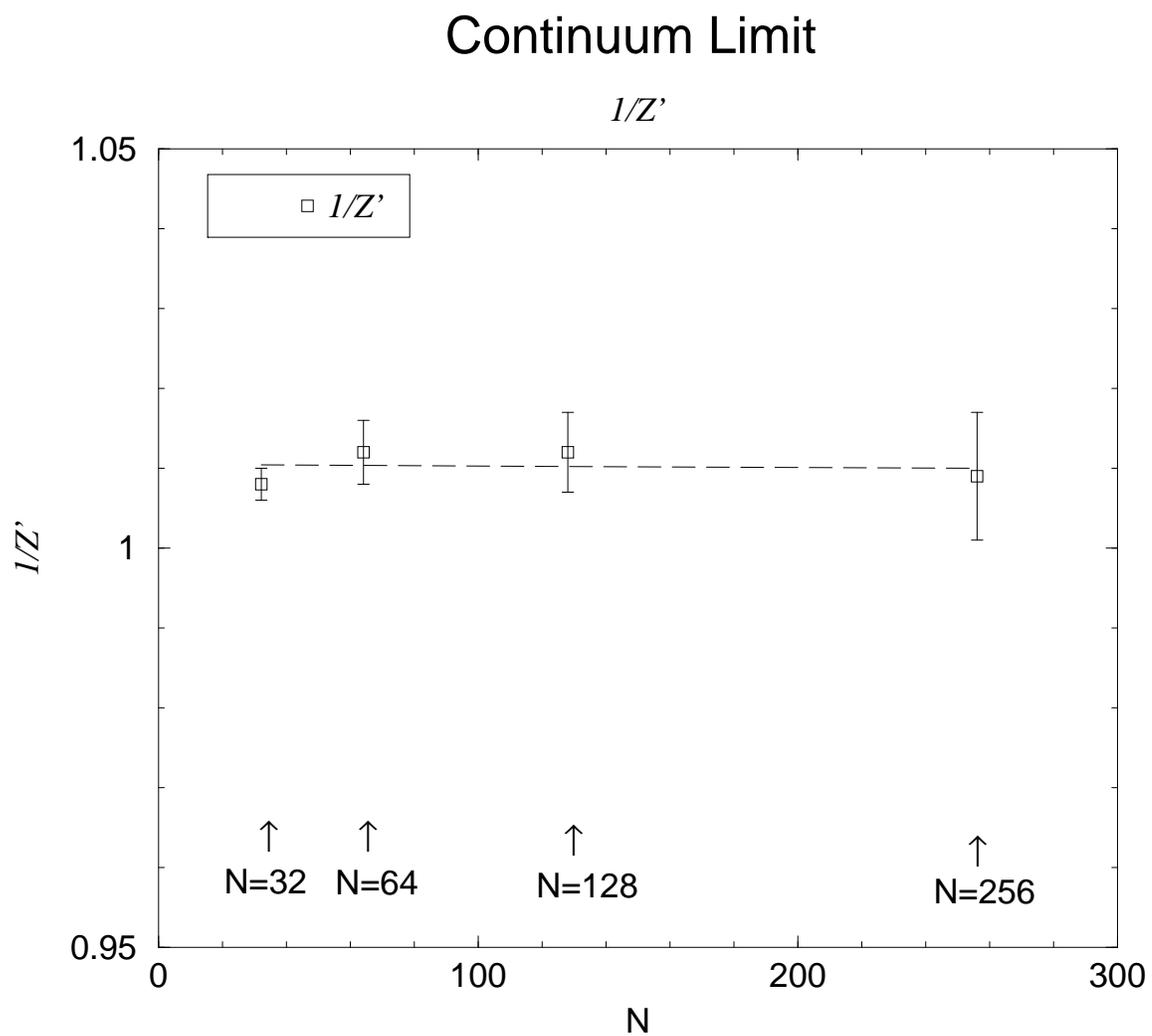}
\vskip 1cm
\caption{Continuum limit for the $\frac1{Z'}$. The
plot shows the insensitivity of lattice size for $\frac1{Z'}$.}
\label{fig:72}
\end{center}
\end{figure}

\begin{figure}[tbp]
\begin{center}
\vskip 1cm
\leavevmode
\epsfxsize 6in
\epsfbox{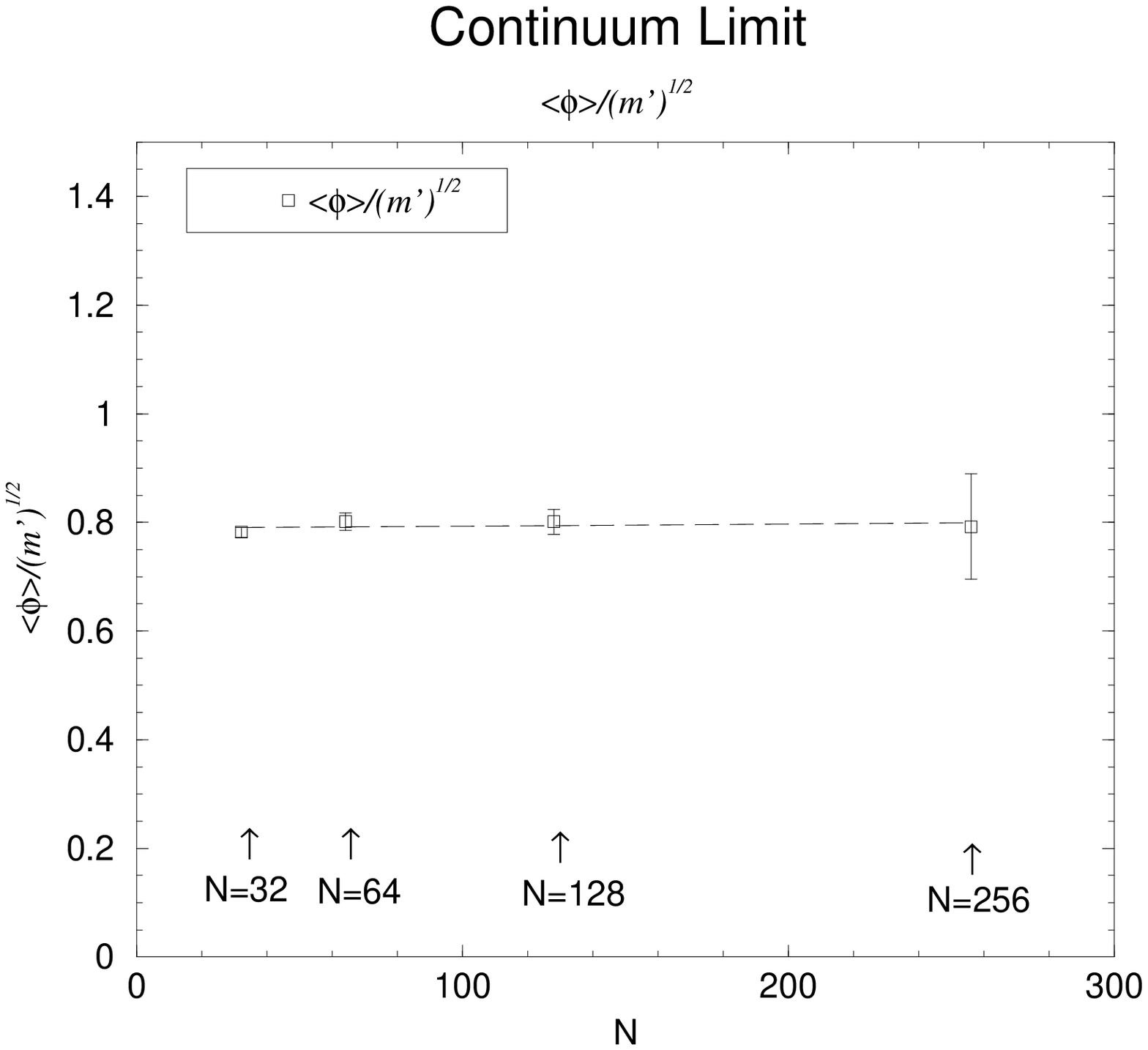}
\vskip 1cm
\caption{Continuum limit for the $\frac{\langle \phi_0 \rangle}{\sqrt{m'}}$. The
plot shows the insensitivity of lattice size for 
$\frac{\langle \phi_0 \rangle}{\sqrt{m'}}$.}
\label{fig:73}
\end{center}
\end{figure}

   The results for the broken symmetry phase are given in Table \ref{tb:71}. 
In going
down from line to line in the Table \ref{tb:71}, we have 
$N \rightarrow 2N$, $\xi_L \rightarrow 2\xi_L$(approximately). This 
corresponds to holding the physical lattice size $L=Na$ fixed while halving
the physical lattice spacing $a$. Here we keep the ratio $L/\xi_L \sim 5.6$. 
This procedure will keep  on a line of ``constant physics" as approach 
to the continuum limit(Figure \ref{figure:phasediag}).
Note that the dimensionless quantities $\lambda /m'$, $\frac1{Z'}-1$, and
$\langle \phi_0 \rangle / \sqrt{m'}$ keep the same values, within errors,
while 
we are halving the lattice space. This shows that these values are 
insensitive to the lattice 
constant $a$ and approaching to the continuum limit.
The Figure \ref{fig:71}, \ref{fig:72}, and \ref{fig:73} plot the 
$\frac{\lambda}{m'}$,  $\frac1{Z'}$ and 
$\frac{\langle \phi_0 \rangle}{\sqrt{m'}}$ in Table \ref{tb:71}. 
The plot shows the insensitivity of lattice size for both those quantities. 
The dashed lines are the fitted lines which show small slopes of the plot. 
However, it is within the error bar.

\begin{table}[tbp]
\caption{The results of our simulations on different size lattices($N$). 
The output correlation length $\xi'_L=1/m'_L$, 
the dimensionless ratios $\lambda /m'$, and  $1/Z'$ are given.}   \label{tb:71}
\vskip 1cm
\begin{center}
\begin{tabular}{|c|c|c|c|c|c|c|}
\hline 
$\mu^2_{0_L}$ & $\lambda_L$ & $N^3$ & $\xi_L$ & $\frac{\lambda}{m'}$ & $\frac1{Z'}$ & $\frac{\langle \phi_0 \rangle}{\sqrt{m'}}$  \\
\hline
-0.161 & 0.2 & $32^3$ & 5.7(1) & 1.13(2) & 1.008(2) & 0.78(1)   \\
-0.078 & 0.1 & $64^3$ & 11.3(4) & 1.13(4) & 1.012(3) & 0.80(2)   \\
-0.0384 & 0.05 & $128^3$ & 22.11(1.0) & 1.11(5) & 1.012(5) & 0.80(2)   \\
-0.01905 & 0.025 & $256^3$ & 45.4(3.0)  & 1.13(8) & 1.010(8) & 0.79(3)   \\

\hline

\end{tabular}
\end{center}
\end{table}

Table \ref{tb:72},\ref{tb:73}, and \ref{tb:74} shows the finite size effect 
of the simulation. We have
$N \rightarrow 2N$, $\xi_L \rightarrow \xi_L$(approximately) while the input
parameters $\mu^2_{0_L}, \lambda_L$ are held fixed. This corresponds to doubling
the physical length of the lattice while keeping the lattice $a$  fixed. 
This check the sensitivity of the output results to lattice size corrections. 
The Tables show small discrepancies at smallest lattice size. 
However,we found that if we keep the ratio $N/\xi_L > 3$, the finite 
size correction can be ignored. In Table \ref{tb:71}, we keep the ratio
$N/\xi_L \sim 5.6$ so that we can ignore the finite size effect while we
are moving along the RG trajectory.

\begin{table}[tbp]
\caption{The finite size effect with fixed input parameters $\mu^2_{0_L}=0.078$,
$\lambda_L=0.1$.} \label{tb:72}
\vskip 1cm
\begin{center}
\begin{tabular}{|c|c|c|c|c|c|c|}
\hline
$\mu^2_{0_L}$ & $\lambda_L$ & $N^3$ & $\xi_L$ & $\frac{\lambda}{m'}$ & $\frac1{Z'}$ & $\frac{\langle \phi_0 \rangle}{\sqrt{m'}}$  \\
\hline
-0.078 & 0.1 & $32^3$ & 12.9(4) & 1.29(4) & 1.0025(10) & 0.73(1)  \\
-0.078 & 0.1 & $64^3$ & 11.3(4) & 1.13(4) & 1.012(4) & 0.80(2)  \\
-0.078 & 0.1 & $128^3$ & 11.2(4) & 1.12(4) & 1.020(5) & 0.80(2)  \\
-0.078 & 0.1 & $256^3$ & 10.8(8) & 1.08(8) & 0.98(2)  & 0.78(3)\\
\hline
\end{tabular}
\end{center}
\end{table}

\begin{table}[tbp]
\caption{The finite size effect with fixed input parameters $\mu^2_{0_L}=0.0384$,
$\lambda_L=0.05$.} \label{tb:73}
\vskip 1cm
\begin{center}
\begin{tabular}{|c|c|c|c|c|c|c|}
\hline
$\mu^2_{0_L}$ & $\lambda_L$ & $N^3$ & $\xi_L$ & $\frac{\lambda}{m'}$ & $\frac1{Z'}$ & $\frac{\langle \phi_0 \rangle}{\sqrt{m'}}$  \\
\hline
-0.0384 & 0.05 & $32^3$ & 21.1(1.6) & 1.06(8) & 1.0002(12) & 0.71(1)  \\
-0.0384 & 0.05 & $64^3$ & 22.1(1.0) & 1.10(5) & 1.003(2) & 0.77(2)  \\
-0.0384 & 0.05 & $128^3$ & 22.1(1.0) & 1.10(5) & 1.012(5) & 0.80(2)  \\
-0.0384 & 0.05 & $256^3$ & 21.9(1.1)& 1.09(6) & 1.015(9)  &  0.79(3)\\
\hline
\end{tabular}
\end{center}
\end{table}

\begin{table}[tbp]
\caption{The finite size effect with fixed input parameters $\mu^2_{0_L}=0.01905$,
$\lambda_L=0.025$.} \label{tb:74}
\vskip 1cm
\begin{center}
\begin{tabular}{|c|c|c|c|c|c|c|}
\hline
$\mu^2_{0_L}$ & $\lambda_L$ & $N^3$ & $\xi_L$ & $\frac{\lambda}{m'}$ & $\frac1{Z'}$ & $\frac{\langle \phi_0 \rangle}{\sqrt{m'}}$  \\
\hline
-0.01905 & 0.025 & $64^3$ & 41.5(6.5) & 1.04(16) & 1.002(3) & 0.78(2)  \\
-0.01905 & 0.025 & $128^3$ & 45.1(3.8) & 1.127(97) & 1.0011(35) & 0.77(2)  \\
-0.01905 & 0.025 & $256^3$ & 45.4(3.0) & 1.13(8) & 1.010(8)  & 0.79(3)  \\
\hline
\end{tabular}
\end{center}
\end{table}

\section{Discussion and the Conclusion}

From Table \ref{tb:71}, on the $256^3$ lattice, we have 
\begin{equation}
\xi_L=45.4 ~>~ \frac{74.24\times m}{\sqrt{3}\lambda} = 37.9, 
\end{equation}
This value satisfies condition \ref{bd:Zp}. From the table, the $1/Z' - 1$ 
is
\begin{equation}
\frac1{Z'} -1 = 0.01 \pm 0.008
\end{equation}
At $95\%$ confedence level, the upper bound of the inclusive cross section is
\begin{equation}
\int ds~\sigma(s)  \le  0.026  \label{result}
\end{equation}
One can see that the numerical result shows no evidence for large 
multiparticle production at high energy in weakly coupled $\phi^4_3$ theory.
To gain further insight into the significance of this results, we 
compare it to the perturbative calculation. 
\begin{equation}
\frac1{Z'} = 1 + \frac{3\lambda}{8\pi m'}\times \frac14 -
\left( \frac{3\lambda}{8\pi m'} \right)^2 \times 0.201214  
\end{equation}
From Table \ref{tb:71}, at the $N^3=256^3$ lattice, we have 
$\frac{\lambda}{m'}=1.13$. So the perturbation result is
\begin{eqnarray}
\frac1{Z'} &=& 1 + 0.1349\times \frac14 - (0.1349)^2 \times 0.20121 \\
           &=& 1 + 0.03372 - 0.00366 \\
           &=& 1.03006
\end{eqnarray}
and the $1/Z'$ in Table \ref{tb:71} shows 
\begin{equation}
\frac1{Z'} = 1.010 \pm 0.008
\end{equation}
we can see that there is little room for multiparticle contributions to
the inclusive cross section.

We can also do a perturbative calculation of 
$\langle \phi_0 \rangle / \sqrt{m'}$ and compare to the lattice Monte Carlo 
result. The result is striking! The perturbation shows
\begin{equation}
\frac{\langle \phi_0 \rangle}{\sqrt{m'}} = \sqrt{\frac{m'}{2\lambda}} \left(
1 + \frac{9\lambda}{4\pi m'}\frac{13}{48} \right)
\end{equation}
For $\frac{\lambda}{m'}=1.13$, the result is
\begin{equation}
\frac{\langle \phi_0 \rangle}{\sqrt{m'}} = 0.665(1+0.219) = 0.811
\end{equation}
The VEV from our simulation($N=256$) is $\langle \phi_0 \rangle = 0.118$ and
from Table \ref{tb:71}, $m'=0.022$. So the simulation results is
\begin{equation}
\frac{\langle \phi_0 \rangle}{\sqrt{m'}} =0.795 ~~\pm ~~0.097
\end{equation}

In conclusion, our results in equation \ref{result} rule out the possibility
of large nonperturbative effects coming from high multiplicity production 
amplitudes in the weakly coupled $\phi^4_3$ theory. The comparison with 
perturbation results from above leaves little room for even small 
nonperturbative effects.

\appendix
\chapter{Multiple zeros case for the function $g^{-1}(z)$}

\begin{figure}[tbp]
\begin{center}
\epsfxsize= 5.0 in
\epsfbox{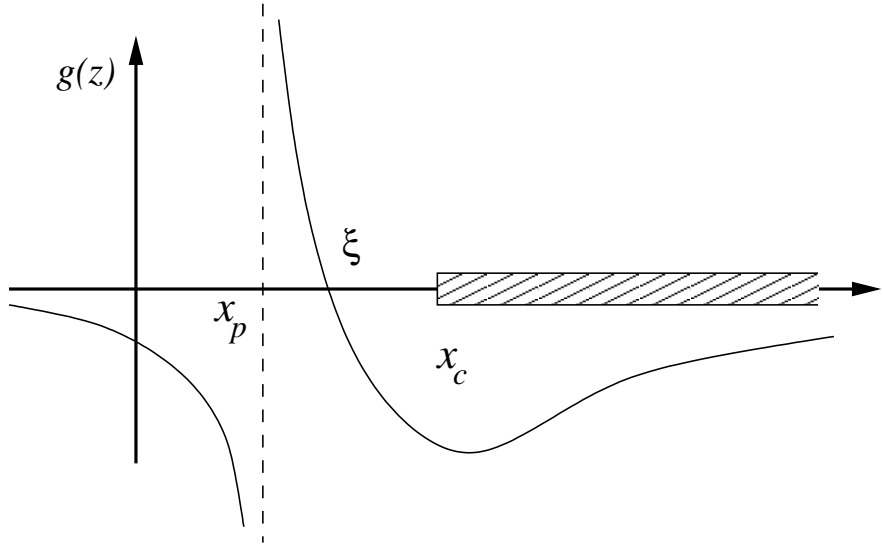}
\caption{The function of $g(z)$ with one zero.}
\label{fig:gzzero}
\end{center}
\end{figure}

In chapter 3, equation \ref{eq:gz-1} leaves open the possibility of zero 
of $g(z)$(pole of $g^{-1}(z)$ 
We deal explicitly with the case of one zero of $g^{-1}(z)$. 
For the multiple zeros case, one only need to replace the single ${\cal C}$ 
term by $\sum {\cal C}_i$. So the equation \ref{eq:gz-1} becomes
\begin{equation}
g^{-1}(z)= f(z)+\frac{{\cal C}}{z-\xi} + \int^{\infty}_{x_c} dx'
\frac{\gamma(x')}{x'-z} \label{eq:gz-1zeros}
\end{equation}
Figure \ref{fig:gzzero} shows the possible function of $g(z)$ with one zero.
note taht ${\cal C}$ is positive.

Because the pole of $g(z)$ is the zero of $g^{-1}(z)$, one can
impose the condition $g^{-1}(x_p)=0$ and $g(z)g^{-1}(z)=1$ into equation
\ref{eq:gz},\ref{eq:gz-1zeros} and get
\begin{equation}
g^{-1}(z)=(z-x_p)\left( 1-\frac{\cal C}{(z-\xi)(\xi-x_p)}+
\int^{\infty}_{x_c}
dx' \frac{\gamma(x')}{(x'-z)(x'-x_p)} \right) \label{gz:invzero}
\end{equation}

In multiparticle production at high energy, $x >> x_c$, equation
\ref{gz:invzero} has a pole on the cut and will give us the results the same
as equation \ref{gm:cr} and \ref{cr:gamm}.

To calculate the renormalization constant $Z$, we  use $g(z)g^{-1}(z)=1$ and
take the limit $z\rightarrow x_p$ to get
\begin{equation}
\frac1Z = 1 + \frac{\cal C}{(\xi - x_p)^2} + \int dx'
\frac{\gamma(x')}{(x'-x_p)^2} + \cdots .\label{Z:invzero}
\end{equation}
After analytic continuation to Euclidean space,  the inverse two point
Green function becomes
\begin{equation}
\frac1{G(p)} = (p^2 +m^2) \left[ 1 + \frac{\cal C}{(p^2 + \xi^2)(\xi^2 - m^2)}
+ \int d\kappa^2 \frac{\gamma(\kappa^2)}{(\kappa^2 - m^2)(\kappa^2 + p^2)}
\right].
\end{equation}
Expand in power of $p^2$
\begin{equation}
\frac1{(p^2 + \xi^2)} = \frac1{\xi^2}\left( 1-\frac{p^2}{\xi^2} +\frac{p^4}{\xi^4} + \cdots \right),
\end{equation}
we have
\begin{eqnarray}
G^{-1}(p) & = & m^2 \left( 1 + \frac{\cal C}{(\xi^2-m^2)\xi^2} +
\int d\kappa^2 \frac{\gamma(\kappa^2)}{(\kappa^2 - m^2)\kappa^2} \right)\nonumber\\
& + & p^2 \left( 1 + \frac{\cal C}{\xi^4} +
\int d\kappa^2 \frac{\gamma(\kappa^2)}{\kappa^4} \right)\nonumber\\
& + & p^4 \left( - \frac{\cal C}{\xi^6}
\int d\kappa^2 \frac{\gamma(\kappa^2)}{\kappa^6} \right) +
\cdots \label{Gi:polyzero}\\
& \equiv & \frac1{Z'} \left(m'^2 + p^2 + (\cdots)p^4 + \cdots \right) \label{tp:invzero}
\end{eqnarray}

From equations \ref{Gi:polyzero} and \ref{tp:invzero}, we have
\begin{equation}
\frac{1}{Z'}  =   1 + \frac{\cal C}{\xi^4} + \int d\kappa^2
~\frac{\gamma(\kappa)}{\kappa^4}  \label{Zp:invzero}
\end{equation}

From equations \ref{eq:Zgamma}  and \ref{Zp:invzero}, the upper
bound of the integrated inclusive cross section becomes
\begin{eqnarray}
\int ds ~\sigma(s) 
& = & Z\int ds \frac{\gamma(s)}{s^2}  \nonumber \\
& = & Z\left( \frac1{Z'} - 1 - \frac{\cal C}{\xi^4} \right) \label{in:cs}  \\
& \leq &  Z\left( \frac1{Z'} - 1 \right) \\
& \leq &  \left( \frac1{Z'} - 1 \right) 
\end{eqnarray}

Because the $\cal C$ is a positive number, the bound remains. It is just
a less effecient bound.


\begin{thebibliography}{99}

\bibitem{Ri:nphb350} A. Ringwald, Nucl. Phys. {\bf B330}, 1(1990)

\bibitem{Es:nphb343} O. Espinosa, Nucl. Phys. {\bf B343},310(1990)

\bibitem{Zj:phyrp70} J. Zinn-Justin, Phys. Rep. {\bf 70}, 109(1981)

\bibitem{Go:phyletb246} H. Goldberg, Phys. Lett. {\bf B246}, 435(1990)

\bibitem{Co:phyletb243} J.M. Cornwall, Phys. Lett. {\bf B243}, 271(1990)

\bibitem{Vo:phyletb293} M.B. Voloshin, Phys. Lett. {\bf B293}, 389(1992)

\bibitem{AKG:nphb341} E.N. Argyres, R.M.P. Kleiss, and C.G. Papadopoulos,
Nucl. Phys. {\bf B341}, 42(1993),{\bf B341}, 57(1993)

\bibitem{Wm:prl} R.D. Mawhinney and R.S. Willey, Phys. Rev. Lett. {\bf 74}, 3728.(1995)

\bibitem{Br:prd46} L.S. Brown, Phys. Rev. {\bf D46}, R4125(1992)

\bibitem{GJ:qphy} J, Glimm and A. Jaffe, {\em Quantum Physics. A Functional
Integral Point of View} (Springer-Verlag, New-York, 1987) 

\bibitem{FF:rwqft} R. Fern\'{a}ndez, Fr\H{o}hlich and A. Sokal,{\em Random
Walks, Critical Phenomena, and Triviality in Quantum Field Theory}
(Springer-Verlag, New York, 1992)

\bibitem{BW:www} see Beowulf web site at {\em http://www.beowulf.org}

\bibitem{Tp:500} see Top500 supercomputer web site at 
{\em http://www.top500.org}

\bibitem{BW:ht} J. Radajewski and D. Eadline {\em Beowulf Howto} and
{\em Beowulf Installation and Administrartion Howto}

\bibitem{Tp:www} Please refer our web site at {\em http://www.phyast.pitt.edu/beowulf}

\bibitem{Tp:500bm} see Top500 cluster  at {\em http://clusters.top500.org}

\bibitem{Vo:itp921} M.B. Voloshin, Minnesota preprint TPI-MINN-92/1-T
\bibitem{Fo:no} N.I. Fuss, Nova Acta Academia Scientiarum Petropolitana {\bf 9}
(1791) 445.

\bibitem{Nu:ci64} F. Lurcat and P. Mazur, {\em Nuovo Cimento} {\bf XXXI}, 
no.1(1964) 140.

\bibitem{PTP:1950} E. Fermi, {\em Progr. Theoret. Phys.}, {\bf 5}, 570(1950). 
see also R. Hagedorn {\em Relativistic Kinematics},(Benjamin Inc, Massachusetts,
1973)

\bibitem{Lt:QFT} I. Montvay and G. Munster, {\em Quantum Fields on a Lattice} 
(Cambridge, United Kingdom 1994), 
or M.E.J. Newman and G.T. Barkema, {\em Monte Carlo Methods in Statistical 
Physics}(Oxford, New York 1999).

\bibitem{SW:cls} R. Swendsen and J.S. Wang, Phys. Rev. Lett. {\bf 58}, 86(1987)

\bibitem{FK:pha} C.M. Fortuin and P.W. Kasteleyn, Physica (Amsterdam) {\bf 57},
536(1972).

\bibitem{Wf:prl23} U. Wolff, Phys. Rev. Lett. {\bf 23}, 361(1989)

\bibitem{KD:mg} D. Kandel, E, Domany, D. Ron, A. Brandt and E. Loh, Jr.,
Phys. Rev. Lett. {\bf 60},1591(1988).

\bibitem{BT:phi4cls} R.C. Brower and P. Tamayo, Phys. Rev. Lett. {\bf 62} 
1087(1989)

\bibitem{HH:dynz} P.C. Hoenberg and B. Halperin, Rev. Mod. Phys {\bf 49}, 435
(1977).

\bibitem{dy:z} The values are taken from Coddington and Bailie(1992), Matz 
{\em et al.}(1994) and Nightingale and Blote(1996). The dynamic exponent of
Metropolis algorithm has not been measured in four dimensions.


\bibitem{tsypin} M. Tsypin, Phys. Rev. Lett. {\bf 73}, 2015(1994); or
M. Tsypin e-print hep-lat/9401034.

\bibitem{Jasnow} J. Rudnick, W. Lay, and D. Jasnow, Phys Rev {\bf E58}, 
2902(1998)



\end{thebibliography}
\end{document}